\documentclass[9pt,twocolumn,twoside]{pnas-new}

\templatetype{pnasresearcharticle} 

\usepackage{url}

\setboolean{displaywatermark}{false}

\def\ben{\begin{equation}}
\def\een{\end{equation}}

\let\l=\lambda   \let\x=\xi

\let\w=\omega

\def\be{\begin{equation}}
\def\ee{\end{equation}}
\def\beq{\begin{equation}}
\def\eeq{\end{equation}}
\def\ba{\begin{array}}
\def\ea{\end{array}}

\def\dalemb#1#2{{\vbox{\hrule height .#2pt
       \hbox{\vrule width.#2pt height#1pt \kern#1pt
               \vrule width.#2pt}
       \hrule height.#2pt}}}

\def\ep{{\epsilon}}
\def\vep{{\varepsilon}}

\title{Theory of the Strange Metal Sr$_3$Ru$_2$O$_7$}

\author[a]{Connie Mousatov}
\author[b]{Erez Berg$^{1,}$}
\author[a,c]{Sean A. Hartnoll}

\affil[a]{Department of Physics, Stanford University, Stanford, CA 94305, USA}
\affil[b]{Department of Condensed Matter Physics, The Weizmann Institute of Science, Rehovot, 76100, Israel}
\affil[c]{Stanford Institute for Materials and Energy Science, SLAC National Accelerator Laboratory, 2575 Sand Hill Road, Menlo Park, CA 94025, USA}

\leadauthor{Mousatov}

\significancestatement{The behavior of `strange metals' has eluded theoretical understanding for some time. It has remained unclear whether strange metals are fundamentally novel forms of matter or whether some unidentified variant of well-understood physics is at work. We show how the behavior of strontium ruthenate, a widely studied strange metal, follows in detail from well-established electronic properties. Specifically, strontium ruthenate contains `hot' electrons that are less quantum mechanical than the other `cold' electrons. A scattering process in which a cold electron becomes hot after colliding with a second cold electron is unusually strong, because the hot electrons are not subject to the Pauli exclusion principle. This fact is seen to underpin the strange metallicity of strontium ruthenate.
}

\correspondingauthor{\textsuperscript{1}To whom correspondence should be addressed. E-mail: erez.berg@weizmann.ac.il}

\begin{abstract}

The bilayer perovskite Sr$_3$Ru$_2$O$_7$ has been widely studied as a canonical strange metal. It exhibits $T$-linear resistivity and a $T\log(1/T)$ electronic specific heat in a field-tuned quantum critical fan. Criticality is known to occur in `hot' Fermi pockets with a high density of states close to the Fermi energy. We show that while these hot pockets occupy a small fraction of the Brillouin zone, they are responsible for the anomalous transport and thermodynamics of the material. Specifically, a scattering process in which two electrons from the large, `cold' Fermi surfaces scatter into one hot and one cold electron renders the ostensibly non-critical cold fermions a marginal Fermi liquid. From this fact the transport and thermodynamic phase diagram is reproduced in detail. Finally, we show that the same scattering mechanism into hot electrons that are instead localized near a two-dimensional van Hove singularity explains the anomalous transport observed in strained Sr$_2$RuO$_4$.  
\end{abstract}

\dates{This manuscript was compiled on \today}

\begin{document}


\maketitle
\thispagestyle{firststyle}
\ifthenelse{\boolean{shortarticle}}{\ifthenelse{\boolean{singlecolumn}}{\abscontentformatted}{\abscontent}}{}

The conventional theory of metallic transport predicts that the approach to the residual resistivity at low temperatures should follow one of several simple power laws. If electron-electron scattering dominates then $\rho \sim T^2$ is expected, with each factor of $T$ coming from the suppression of final electron states due to Pauli exclusion. Dominant electron-phonon scattering instead leads to $\rho \sim T^5$ in three dimensions. `Strange metals' can be defined as metals exhibiting an anomalous temperature dependence of the low temperature resistivity. The $T$-linear behavior $\rho \sim T$ has been widely observed, but other scalings including $\rho \sim T^{3/2}$ and $\rho \sim T^2 \log T$ have also been reported. These are all stronger than the conventional $T^2$ scaling due to electronic scattering. A common scenario is that the anomalous scaling is observed above a `transport temperature' $T_\text{tr}$, and that $T_\text{tr} \to 0$ at some point in the phase diagram, often associated with quantum criticality \cite{sachdevkeimer}.

A compelling theory of a strange metal must connect the anomalous transport behavior to other measurements, including thermodynamic and spectroscopic probes. Given the similarity of the strange metal phenomonon across materials, a thorough grounding of anomalous transport in specific microscopic electronic properties of a single material has the potential to uncover mechanisms that may be at work more broadly. Here we show how such a comprehensive theory can be achieved for the bilayer perovskite Sr$_3$Ru$_2$O$_7$. This material has been widely studied as a prototypical strange metal \cite{Grigera329}, with $T$-linear resistivity in a quantum critical fan and diverging effective electron masses as the critical field is approached at low temperature.

The key fact will be that Sr$_3$Ru$_2$O$_7$ is known from photoemission measurements to exhibit very shallow, small Fermi pockets. These lead to a sharp peak in the density of states close to the chemical potential \cite{PhysRevLett.101.026407}. The distance of the peak from the chemical potential is comparable to the scale $T_\text{tr}$ above which strange metal transport is observed at zero field. Furthermore, it is the `hot' electrons in these pockets that go critical as the field is tuned. These facts suggest that the `hot' (h) electron pockets should be responsible for the anomalous transport, but a suitable mechanism has been lacking. The hot electrons are irrelevant as carriers of charge, because they are short circuited by the faster `cold' (c) electrons \cite{PhysRevB.51.9253, PhysRevLett.82.4280}. Instead, a distinctive cc $\to$ ch scattering process, in which one cold electron is scattered by another into a hot pocket, leads to strange metal transport by the cold electrons. As previously noted in the context of cuprates \cite{varma1, varma2}, electrons in a peak close to the chemical potential are classical above a low energy scale, so there is no phase space suppression for scattering into the hot electron pocket and hence $\rho \sim T$.

In addition to the $T$-linear resistivity \cite{Grigera329,PhysRevLett.88.076602,Bruin804}, cc $\to$ ch scattering above the temperature $T_\text{tr}$ produces the observed anomalous specific heat \cite{Rost16549} and optical conductivity \cite{PhysRevB.78.155132} of Sr$_3$Ru$_2$O$_7$. At temperatures below $T_\text{tr}$, the dominance of cc $\to$ ch scattering as the critical field is approached leads to a violation of the Kadowaki-Woods relation $A \sim \gamma^2$ between the resistivity $\rho \sim A \, T^2$ and specific heat coefficient $\gamma\equiv c/T$. Fig. \ref{fig:A_vs_gamma} below shows that instead $A \sim \Delta \gamma$ for Sr$_3$Ru$_2$O$_7$ \cite{Grigera329, Rost1360,Rost16549}, as our model predicts. Here $\Delta \gamma$ is the enhancement of the specific heat as the critical field is approached.

The scattering mechanism we have identified may be relevant in other contexts. We will show that the same cc $\to$ ch scattering  explains the anomalous $\rho \sim T^2 \log T$ of a different ruthenate, Sr$_2$RuO$_4$, as it crosses a Lifshitz transition \cite{PhysRevLett.120.076602}. Beyond ruthenates, in many families of strange metals strong quantum critical scattering is only directly experienced by a relatively small fraction of the electrons, localized in `hot regions' or in `hot bands'. The conceptual understanding of transport Sr$_3$Ru$_2$O$_7$ developed here gives a solid starting point for unravelling the mystery of strange metals more broadly.

\subsection{Quantum criticality in Sr$_3$Ru$_2$O$_7$}

Criticality in Sr$_3$Ru$_2$O$_7$ is field-tuned, with a critical $c$ axis field of $H_c \approx 7.9 \text{T}$ \cite{Grigera329}, and displays 
\begin{figure}[h]
\begin{center}
\includegraphics[width=0.4\textwidth]{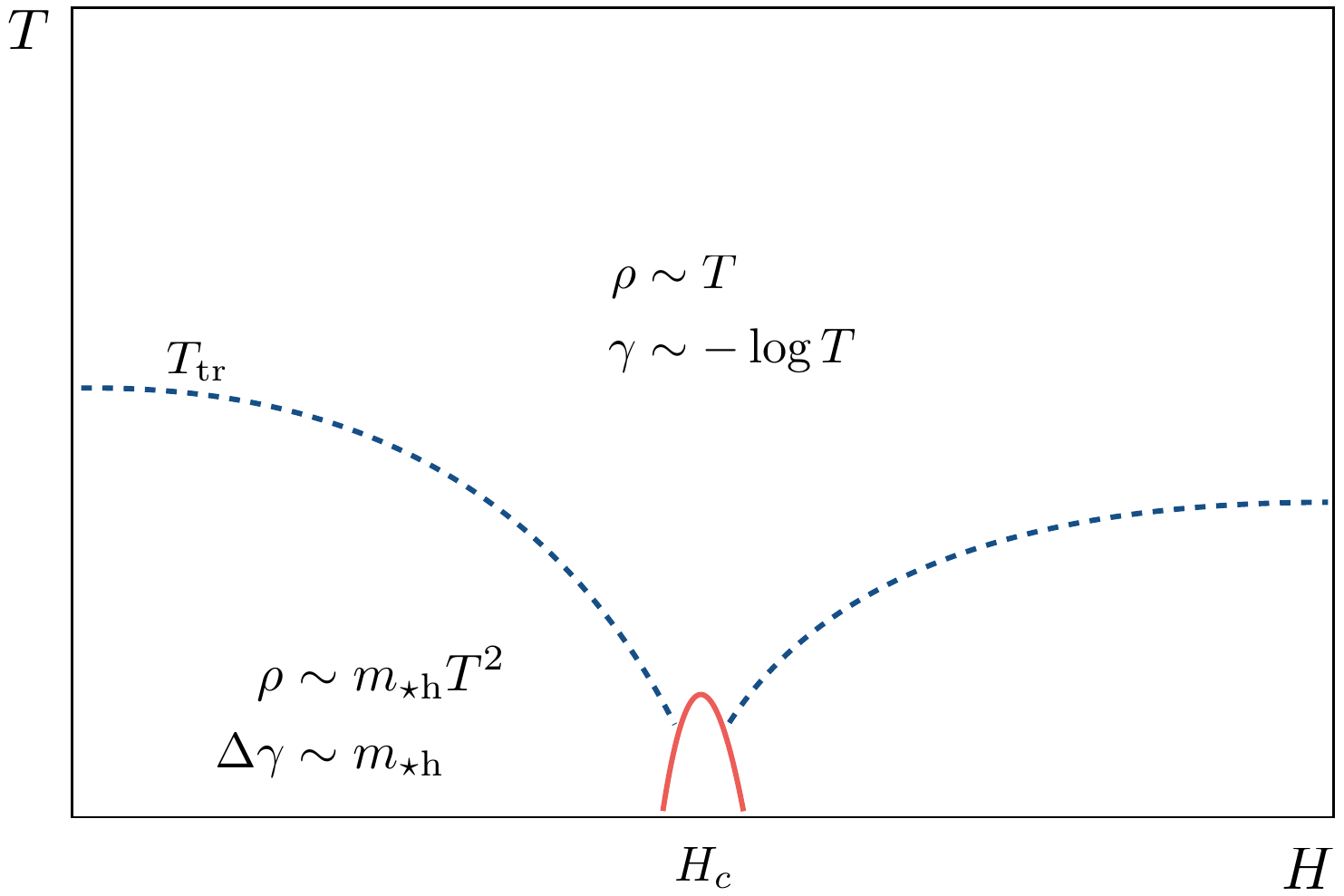}
\caption{Schematic phase diagram of Sr$_3$Ru$_2$O$_7$ as a function of field and temperature. There is a marginal Fermi liquid of cold electrons above $T_\text{tr}$. In the conventional Fermi liquid below $T_\text{tr}$, transport and specific heat are dominated by the diverging hot electron mass $m_{\star\text{h}}$ as $H \to H_c$.}
\label{fig:phase}
\end{center}
\end{figure}
several canonical features of metallic quantum criticality.
The transport and thermodynamic phase diagram is sketched in Fig. \ref{fig:phase}. In more detail:
\begin{enumerate}
\item At $H=0$ the resistivity is $T$-linear above $T_\text{tr}\sim20 \text{K}$, and goes as $T^2$ at the lowest temperatures. As $H \to H_c$, the scale $T_\text{tr}$ apparently collapses towards zero \cite{Grigera329,PhysRevLett.88.076602,Bruin804}. \label{lab:1}

\item As $H \to H_c$ at low temperatures, an effective mass $m_\star$ is strongly enhanced. The growth of $m_\star$ is seen in the coefficient $A$ of the low temperature resistivity $\rho \sim A T^2$ \cite{Grigera329}, NMR spin relaxation \cite{PhysRevLett.95.127001} and the low temperature specific heat coefficient $\gamma \equiv c/T$ \cite{Rost1360,Rost16549}. However, we see in Fig. \ref{fig:A_vs_gamma} that the specific heat enhancement $\Delta \gamma$ follows the scaling $A \sim \Delta \gamma$, distinct from the conventional Kadowaki-Woods relation $A \sim \gamma^2$.\label{lab:2}

\begin{figure}[h] 
\begin{center}
\includegraphics[width=0.35\textwidth]{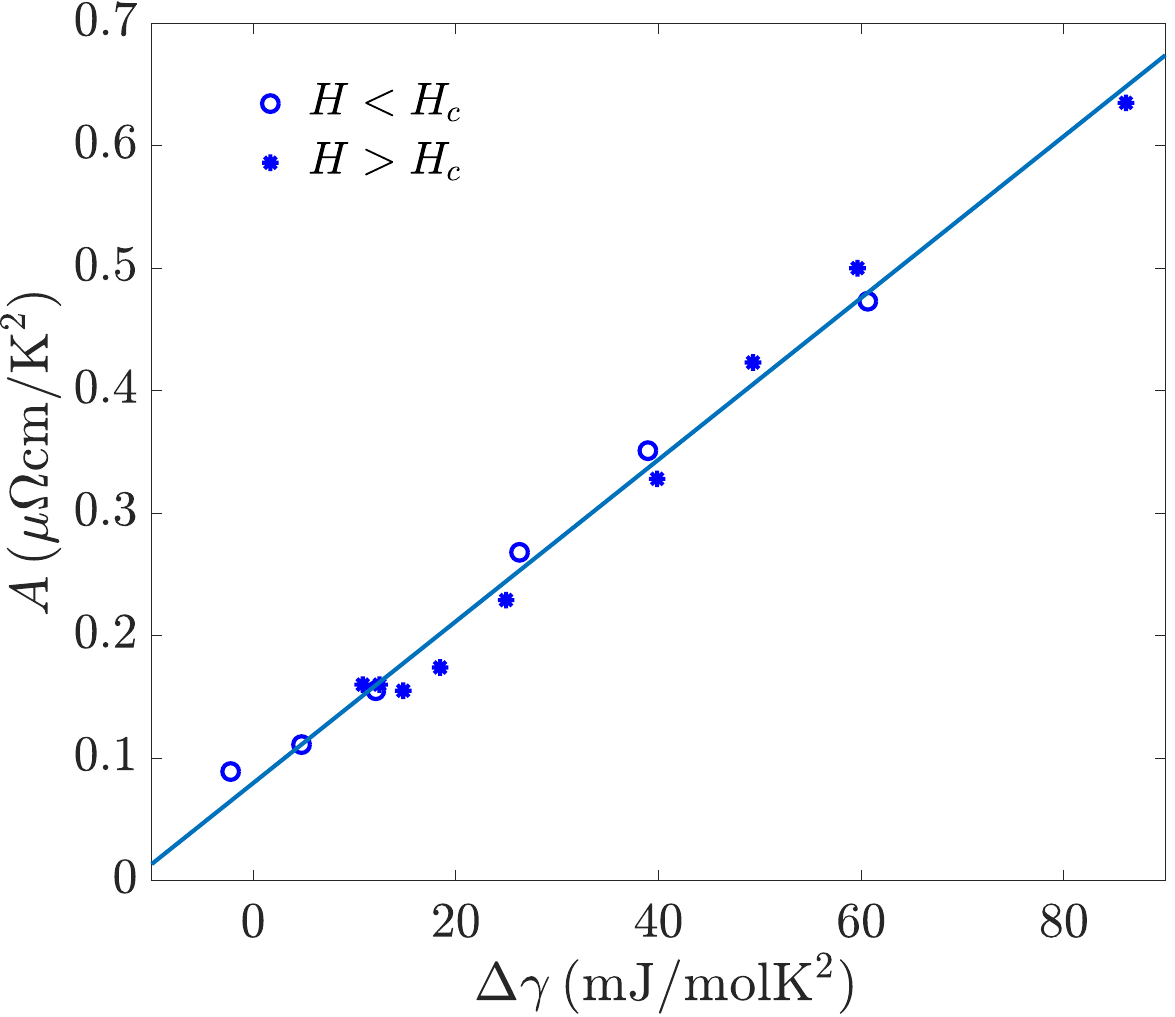}
\caption{As $H \to H_c$ at low temperatures, the coefficient $A$ of the $T^2$ resistivity increases by a factor of $\sim 7$ \cite{Grigera329}, and is linearly proportional to the specific heat enhancement $\Delta \gamma$ \cite{Rost1360,Rost16549}.}
\label{fig:A_vs_gamma}
\end{center}
\end{figure}

\item At $H=H_c$ a low temperature divergence $\gamma \sim \log T$ is observed \cite{Rost16549, PhysRevB.97.115101}, until cut off at a spin ordering transition at a temperature of about $1$K \cite{Lester2015}.\label{lab:4a}

\item An extended Drude fit to $\sigma(\omega)$ in the $T$-linear regime reveals a scattering rate $\Gamma(\omega) \sim \omega$ for $\omega \gtrsim T$ \cite{PhysRevB.78.155132}.\label{lab:5}

\end{enumerate}

\subsection{Hot pockets and cc $\to$ ch scattering}

Quantum criticality in Sr$_3$Ru$_2$O$_7$ involves a dichotomy between cold and hot electrons. The hot electrons have a large density of states already in zero field, and only the hot electrons go critical as a function of field. Specifically:

\begin{enumerate}

\item Angle-resolved photoemission (ARPES) reveals a complicated fermiology with multiple bands \cite{PhysRevLett.101.026407, AllanARPES}. Quantum oscillations show that all except one set of bands --- the $\gamma_2$ pockets, which are not seen in Shubnikov-de Haas oscillations --- do not exhibit a strongly divergent mass at the critical field \cite{PhysRevB.81.235103}. The $\gamma_2$ pockets are remarkably shallow, and give rise to a peak in the density of states at a few meV from the chemical potential at zero field \cite{PhysRevLett.101.026407, AllanARPES}.\label{lab:3}

\item The coefficient $\gamma$ as a function of temperature shows a maximum at $T_\text{pk} \approx 7 \text{K}$ at zero field \cite{Rost16549}. A similar peak is seen at $T_\text{pk}' \approx 17 \text{K}$ in the magnetic susceptibility \cite{CAVA1995141, PhysRevB.62.R6089}. As $H \to H_c$, $T_\text{pk}$  collapses to zero.\label{lab:4b}

\end{enumerate}

At zero field the hot electron peak in the density of states seen in photoemission quantitatively explains the peaks seen in the specific heat coefficient $\gamma$ and magnetic susceptibility $\chi$. Using the measured density of states of the $\gamma_2$ pockets \cite{PhysRevLett.101.026407} in standard free electron formulae for $\gamma$ and $\chi$ leads to peaks at temperatures of $9$K and $19$K, respectively. These are in good agreement with the zero field experimental values mentioned above. This is the first imprint of the hot electrons on physical observables. In the remainder we show that cc $\to$ ch scattering into the hot electron pockets fundamentally determines the transport behavior both above and below $T_\text{tr}$.

\begin{figure}[h]
\begin{center}
\includegraphics[width=0.45\textwidth]{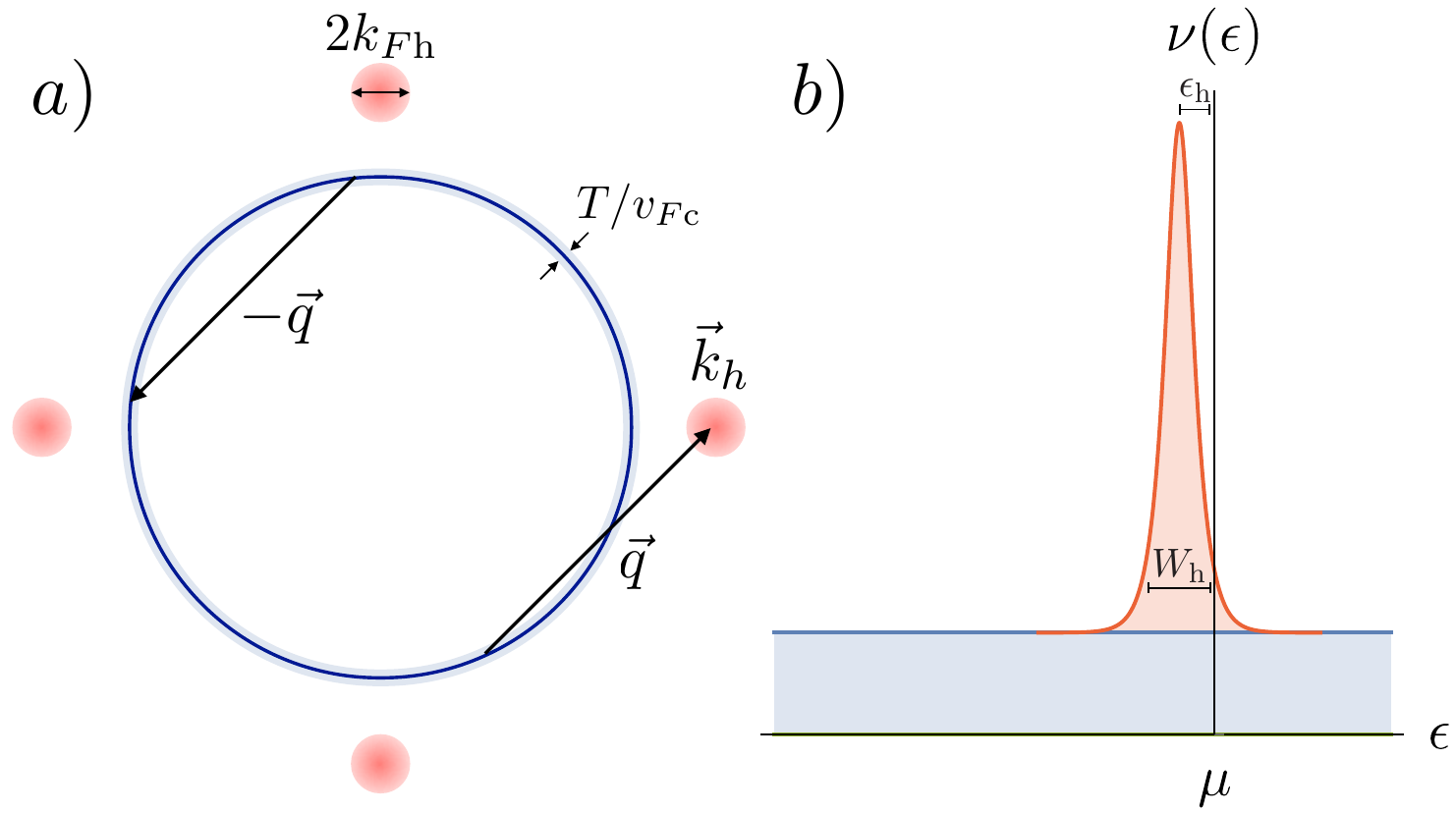}
\caption{Left: Illustration of cc $\to$ ch scattering, in which one of the fermions is scattered from the cold Fermi surface (blue) into the hot region (red). In general, the hot regions could also be on the same Fermi sheet as the cold fermions. Right: a sharp peak in the density of states $\nu(\epsilon)$ close to the chemical potential, due to the hot fermions.}
\label{fig:hot_cold}
\end{center}
\end{figure}

Fig. \ref{fig:hot_cold} illustrates the interplay of hot and cold electrons. The Fermi surface mostly consists of regions with a high Fermi velocity, as well as a set of `hot pockets' with a large density of states peaked near or at the chemical potential. The essential feature for the present discussion is that the high density of states is localized in a small part of the Brillouin zone, whereas the light portions of the Fermi surface are much more extended.

The density of states of the `hot electrons' is high, whereas the `cold electrons' away from the hot spots dominate the current due to their higher Fermi velocity. It is therefore natural for the transport lifetime to be determined by scattering processes where both cold and hot electrons are involved. Several such processes are possible: a cold electron can scatter off a hot one, denoted ch $\rightarrow$ ch; a pair of cold electrons can scatter into a cold and a hot electron (cc $\rightarrow$ ch); etc.

The ch $\rightarrow$ cc or cc $\rightarrow$ ch processes involve large momentum transfer, and contribute significantly to transport. Moreover, for a large enough cold Fermi surface, all cold electrons can participate in cc $\rightarrow$ ch processes and no `short circuiting' occurs. Other types of scattering are either small angle, and do not efficiently degrade current, or are not active across the entire Fermi surface. We discuss this further in the Supplementary Information (SI). In the following we show that cc $\rightarrow$ ch scattering reproduces the phase diagram in Fig. \ref{fig:phase}.

\subsection{A Marginal Fermi Liquid from classical fermions}
\label{sec:zerofield}

The peak in the density of states due to the `hot' $\gamma_2$ pockets (at $H=0$, from ARPES) occurs at a distance from the chemical potential that roughly coincides with the temperature above which $T$-linear resistivity is observed. Let the distance of the peak from the chemical potential be $\epsilon_\text{h}$, and its width be $W_\text{h}$ (see Fig.~\ref{fig:hot_cold}). At zero field, $W_h \lesssim |\epsilon_{\text{h}}|$. At temperatures above $|\epsilon_\text{h}| + W_\text{h}$ the $\gamma_2$ fermions are classical and non-degenerate. A cc $\to$ ch scattering process can therefore be expected to be suppressed by $T$ rather than $T^2$. We now verify that this is the case by showing that this scattering leads to Marginal Fermi Liquid (MFL) \cite{PhysRevLett.63.1996} behavior of the cold fermions, as has been previously noted \cite{varma1, varma2}.

To connect with the MFL cleanly, we write the scattering process as the interaction of a cold fermion with a bosonic mode generated by the cold-hot fermion polarizability. We treat the interaction perturbatively, starting from free cold fermions with spectral weight $\text{Im} \, G^R_\text{c}(\omega,\vec k) = \pi \delta(\omega - \ep_{\text{c},\vec{k}})$. Due to the interaction, 
the cold fermions acquire the self energy (we work in $d$ spatial dimensions for generality)
\begin{eqnarray}
\lefteqn{\Sigma''_\text{c}(\omega,\vec{k}) =  \lambda^2 \int \frac{d^d q}{(2 \pi)^d} \int \frac{d \Omega}{\pi} \frac{f(\omega-\Omega) b(\Omega)}{f(\omega)}} \nonumber \\ 
& & \times \text{Im} \, G^R_\text{c}(\omega - \Omega,\vec{k}-\vec{q}\,) \, \Pi''_\text{ch}(\Omega,\vec{q}\,) \,. \label{eq:Sigma}
\end{eqnarray}
In this form, the above equation is transparently Fermi's Golden rule: $\lambda$ is the four-fermion coupling constant and the Fermi-Dirac and Bose-Einstein distributions are $f(\Omega) = 1/(e^{\Omega/T} + 1)$ and $b(\Omega) = 1/(e^{\Omega/T} - 1)$. In textbooks~\cite{Mahan2000many}, [\ref{eq:Sigma}] is commonly expressed differently, using the identity $f(\omega-\Omega) b(\Omega)/f(\omega) = f(\Omega-\omega) + b(\Omega)$. The polarizability is
\begin{eqnarray}
\label{eq:Pi}
\lefteqn{\Pi''_\text{ch}(\Omega,\vec{q}\,) = \int \frac{d^d k'}{(2 \pi)^d} \int \frac{d \Omega'}{\pi} \frac{f(\Omega'+\Omega) f(-\Omega')}{b(\Omega)}} \nonumber \\ 
& & \times \text{Im} \, G^R_\text{c}(\Omega',\vec{k}') \, \text{Im} \, G^R_\text{h}(\Omega'+\Omega,\vec{k}'+\vec{q}\,) \,.
\end{eqnarray}
Here $\text{Im} \, G^R_\text{h}(\Omega'+\Omega,\vec{k}'+\vec{q}\,)$ is the spectral weight of the hot fermions, which we can keep general. Similarly to above, we have written equation [\ref{eq:Pi}] in a manifestly physical way, related to common formulations though the identify
$f(\Omega'+\Omega) f(-\Omega')/b(\Omega) = f(\Omega') - f(\Omega + \Omega')$.

At temperatures greater than 
$|\epsilon_\text{h}| + W_\text{h}$,
we have 
$T \sim |\omega| \sim |\Omega| \sim |\Omega'| \gg |\Omega'+\Omega|$ wherever $\text{Im} \, G^R_\text{h}(\Omega'+\Omega,\vec{k}'+\vec{q}\,)$ is nonzero. This leads to simplifications in [\ref{eq:Pi}]. The cold fermions are at energies $T \ll v_{F\text{c}} k_{F\text{c}} = E_{F\text{c}}$ (where $k_{F\text{c}}$, $v_{F\text{c}}$, $E_{F\text{c}}$ denote the cold fermions' Fermi momentum, Fermi velocity, and Fermi energy, respectively), which allows the cold fermion dispersion to be linearized about the Fermi energy. In this way we obtain the explicitly MFL form
\be
\Pi''_\text{ch}(\Omega,\vec{q}\,) \approx  F(\vec{q}\,) \tanh \frac{\Omega}{2T}  \,.\label{eq:MFL}
\ee
Taking the cold Fermi surface to be round for simplicity,
\be
F(\vec{q}\,) \approx \frac{\pi}{2} \int \frac{d^d k'}{(2 \pi)^d}  \delta(|\vec k'| - k_{F\text{c}}) \int \frac{d \Omega'}{\pi}  \frac{1}{v_{F\text{c}}} \text{Im} \, G^R_\text{h}(\Omega'+\Omega, \vec{k}'+\vec{q}\,) \,. \label{eq:Fq}
\ee
Equation [\ref{eq:MFL}] was noted in Ref.~\cite{varma1}. It is then well known that inserting the MFL polarization [\ref{eq:MFL}] into the expression [\ref{eq:Sigma}] for the self-energy leads to $\Sigma_\text{c}'' \sim \max(T,\omega)$ \cite{PhysRevLett.63.1996}. Thus, there is a $T$-linear scattering rate that becomes $\omega$-linear at higher frequencies. This explains the experimental resistivity and optical conductivity noted above. A careful evaluation of the cc $\to$ ch scattering rate in the SI, using the experimentally determined density of states in the $\gamma_2$ pockets \cite{PhysRevLett.101.026407} to evaluate [\ref{eq:Pi}], shows that $T$-linear scattering is found above precisely the observed $T_\text{tr} \approx 20 K$. Because cc $\to$ ch scattering is large angle, this result for the single particle scattering rate also determines the transport lifetime.

The result [\ref{eq:MFL}] is obtained for any form of the hot fermion Green's function. The only assumption is that the temperature is above the characteristic energy scales of the hot fermions 
(such as $W_\text{h}$ and $\epsilon_\text{h}$). We can estimate the magnitude of the scattering rate from [\ref{eq:Fq}] by taking the hot band to be flat, so that $\text{Im} \, G^R_\text{h}(\w,\vec{k}) = \pi \delta(\omega)$ for some region of area/volume $k_{F\text{h}}^d$ in the Brillouin zone. The hot regions are smaller than the cold Fermi sea, so that $k_{F\text{h}} \ll k_{F\text{c}}$. In this case, the phase space that the cold fermion can scatter into is constrained by the size of the hot regions; evaluating the integrals one finds the scattering rate $\Gamma_\text{c} \equiv \Sigma_\text{c}''(\omega=0) \sim \left[\lambda^2 k_{F\text{c}}^{d-2}k_{F\text{h}}^{d}/v_{F\text{c}}^2\right] T$. We give some details of the kinematics leading to this result in the SI. Specializing to $d=2$ and restoring units:
\be\label{eq:Gamma1}
\Gamma_\text{c} \sim \lambda^2 \frac{k_{F\text{h}}^{2}}{v_{F\text{c}}^2} \frac{k_B T}{\hbar} \,.
\ee
For small $k_{F\text{h}}$, the prefactor of $k_B T/\hbar$ in [\ref{eq:Gamma1}] is small and the perturbative computation is controlled. For Sr$_3$Ru$_2$O$_7$ we can estimate the prefactor using numbers from \cite{Bruin804}. There are eight hot pockets with $\sum_i k_{F\text{h},i}^2 \approx 0.03 \text{A}^{-2}$. The inverse velocity of the cold Fermi surfaces in [\ref{eq:Gamma1}] controls the available phase space for scattering. Averaging over bands we estimate $\langle v_{F\text{c}}^{-1} \rangle \approx 0.5 \times 10^{-4} \text{s/m}$. The microscopic coupling $\lambda$ has units of $\hbar$/mass. Again averaging over bands we estimate $\lambda \sim \langle v_{F\text{c}}/k_{F\text{c}}\rangle \approx 0.1 \hbar/m_\text{e}$. These numbers lead to $\Gamma \sim k_B T/\hbar$, consistent with the observed `Planckian' scattering rate \cite{Bruin804, Zaanen2004}.

A similar computation can be done for the ch $\to$ ch process, in which a cold electron scatters off an electron in the hot region. Both the initial and final states of the hot electron have no Fermi-Dirac suppression. One finds $\Sigma''_\text{c}(\omega,\vec{k}) \sim \lambda^2 k_{F\text{h}}^2 \int \frac{d^2q}{(2\pi)^2} \delta(\omega - \epsilon_{\text{c},\vec{k}-\vec{q}})$. This scattering therefore behaves as elastic disorder, leading to a scattering rate proportional to $\lambda^2 k_{F\text{h}}^3/v_{F\text{c}}$ (the $q$ integral is restricted to a region of the same size as the hot region). It contributes a temperature-independent term to resistivity. However, because $k_{F\text{h}} \ll k_{F\text{c}}$, this is necessarily small angle scattering for the cold electrons. Therefore, the effective transport scattering rate is suppressed by an additional factor of $(k_{F\text{h}}/k_{F\text{c}})^2$, so that $\Gamma_\text{tr} \sim k_{F\text{h}}^4/k_{F\text{c}}^4 \times k_{F\text{h}} v_{F\text{c}}/\hbar$ (assuming for simplicity that $\lambda \sim v_{F\text{c}}/k_{F\text{c}}$ is set by a single cold Fermi surface).

There is also a conventional $T^2$ contribution to the resistivity from cc $\to$ cc scattering. For the MFL $T$-linear scattering to dominate, the phase space for scattering into cold fermions must be smaller than that for scattering into hot fermions: $k_{F\text{c}} k_B T/v_{F\text{c}} \ll k_{F\text{h}}^2$. Geometrically, cc $\to$ ch scattering dominates when the area of the hot region is greater than the area of the thermally broadened cold annulus shown in Fig. \ref{fig:hot_cold}. This requirement, that $k_B T \ll k_{F\text{h}}^2/k_{F\text{c}}^2 \times E_{F\text{c}}$, can be satisfied despite $k_{F\text{h}}$ being small because $E_{F\text{c}}$ is a high energy scale. In the SI we verify that the phase space for cc $\to$ ch scattering indeed dominates in the $T$-linear regime of Sr$_3$Ru$_2$O$_7$.

\subsection{Low temperature approach to the critical field}
\label{sec:lowT}

The maximum in the density of states in Fig.~\ref{fig:hot_cold} is at a distance $|\epsilon_\text{h}|$ from the chemical potential. At temperatures well below this scale all of the fermions are degenerate, and electron-electron scattering leads to the usual $T^2$ resistivity. If $|\epsilon_\text{h}|<W_\text{h}$, however, there remains a degenerate Fermi pocket of hot fermions that makes a large contribution to the density of states and cc $\to$ ch scattering is still important. In this regime $\omega \sim \Omega \sim \Omega' \sim \Omega+\Omega'\sim T$, and the formulae above give the scattering rate (re-instating factors of $k_B$ and $\hbar$ in the final term)
\be\label{eq:gammalow}
\Gamma_\text{c} \sim \l^2 \frac{k_{F\text{h}}}{v_{F\text{h}} v_{F\text{c}}^2}  T^2 \sim \frac{m_{\star\text{h}}}{m_{\star\text{c}}}  \frac{(k_B T)^2}{\hbar E_{F\text{c}}} \,,
\ee
where we define the mass of the heavy electrons through their density of states at the Fermi level: $m_{\star\text{h}}=\nu_\text{h}(0)/2\pi$, and similarly for $m_{\star\text{c}}$. In the final term in [\ref{eq:gammalow}] we again assumed for simplicity that $\lambda \sim \hbar/m_{\star\text{c}}$ is set by a single cold Fermi surface.
See the SI for details of the kinematics leading to [\ref{eq:gammalow}].
The important factor is the ratio $m_{\star\text{h}}/m_{\star\text{c}}$ that, we will posit, becomes large as the critical field is approached. The scattering rate $\Gamma_\text{c}$ determines the resistivity through the Drude formula $\rho = m_{\star\text{c}} \Gamma_\text{c}/(n_\text{c} e^2)$. The resulting scaling of the resistivity with the large hot mass is therefore $\rho \propto m_{\star\text{h}} T^2$. In contrast,
if all fermions have a strongly enhanced mass (become hot) or if ch $\to$ ch scattering dominates (for example, if the hot band is large in the Brillouin zone), then $\rho \propto m_{\star\text{h}}^2 T^2$.

At the same low temperatures, the electronic specific heat is also given by the conventional Fermi liquid formula. If $m_{\star\text{h}} \gg m_{\star\text{c}}$ then hot electrons dominate the specific heat, even while cold electrons dominate transport. The enhancement of the specific heat coefficient $\gamma$ due to the increasing mass of the hot fermions is $\Delta \gamma \sim k_B^2 m_{\star\text{h}}$. Together with the results in the previous paragraph, this leads to $A \sim \Delta \gamma$, where $A$ is the coefficient in the resistivity $\rho \sim A T^2$. The notation $\Delta \gamma$ refers to the fact that the specific heat of the cold fermions has been subtracted out.\footnote{As $m_{\star\mathrm{h}}$ diverges, $m_{\star\mathrm{c}}$ also becomes enhanced due to cc$\rightarrow$ch scattering. We will see shortly that the enhancement is only logarithmic: $\Delta m_{\star\mathrm{c}} \propto k_{F\text{h}}^2 \log\left(m_{\star\mathrm{h}}v_{F\text{c}}/k_{F\text{h}}\right)$. Close to the critical field this growth is parametrically weaker than that of $m_{\star\mathrm{h}}$, plausibly explaining why it is not seen in quantum oscillations \cite{PhysRevB.81.235103}. That said, our interpretation of the specific heat data in the SI suggests that $\Delta m_{\star\mathrm{c}} \sim m_{\star\mathrm{c}}$ is not negligible. \label{foot:mstar}}
We have recovered precisely the observed linear scaling shown in Fig. \ref{fig:A_vs_gamma}. The key input is the hypothesis that $m_{\star\text{h}}$ becomes large as $H \to H_c$, which leads to the dominance of cc $\to$ ch scattering. This scaling is distinct from the Kadowaki-Woods relation $A \sim \gamma^2$ \cite{KADOWAKI1986507}, which instead follows from the resistivity $\rho \propto m_{\star\text{h}}^2 T^2$.

Microscopically, the mass enhancement has been assumed to be tied up with metamagnetic quantum criticality of the hot electrons \cite{PhysRevLett.86.2661,Grigera329,PhysRevLett.88.217204, PhysRevLett.116.226402}. The results above are not sensitive to the details of this physics beyond requiring a divergent $m_{\star\text{h}}$ in a small region of the Brillouin zone. However, in using [\ref{eq:Fq}] we are also assuming that the integrated single-particle spectral weight of the hot electrons remains finite as their mass diverges. This leads to a picture in which the growth of the mass is primarily a band structure effect associated with a strongly enhanced density of states at the Fermi energy as $H \to H_c$ \cite{toapp}. This spectral weight must survive the presence of any quantum critical scattering.

\subsection{The collapse of $T_\text{tr}$ at the critical field}

The hot electrons dominate the specific heat at low temperatures, especially as the critical field is approached. At temperatures above the peak in the density of states their contribution becomes small: $\gamma_\text{hot} \sim k_{F\text{h}}^2 \max\left(W_\text{h}^2,\epsilon_\text{h}^2\right)/(k_B T^3)$. In computing $\gamma_\text{hot}$ here the chemical potential is kept fixed and constant, controlled by the large total number of cold electrons. The cold fermions dominate the specific heat in this regime. The MFL scattering rate $\Sigma_\text{c}'' \sim \max(T,\omega)$ implies a logarithmic effective mass enhancement so that $\Delta \gamma_\text{cold} \sim k_B^2 \Delta m_{\star\text{c}} \sim k_B^2 m_{\star\text{c}} \cdot k_{F\text{h}}^2/k_{F\text{c}}^2 \cdot \log (\Lambda/T)$. Here $m_{\star\text{c}}$ is the unrenormalized cold electron mass. The energy cutoff scale $\Lambda \sim v_{F\text{c}} k_{F\text{h}}$ because MFL scattering $\Gamma_\text{c} \sim \omega$ requires $\omega/v_{F\text{c}} \lesssim k_{F\text{h}}$, in order for the scattering phase space to be determined by the hot fermion pocket (see SI). Precisely such a logarithmic temperature dependence is seen cleanly at low temperatures at the critical field, as we recalled above. The onset of logarithmic behavior is also visible in the data away from the critical field, at temperatures above $T_\text{tr}$.
At temperatures below $T_\text{tr}$, within the hot electron peak in the density of states, the hot fermions become degenerate and the cold fermion mass enhancement is cut off as $\gamma_\text{cold} \sim k_B^2 m_{\star\text{c}} \cdot k_{F\text{h}}^2/k_{F\text{c}}^2 \cdot \log \Big(\Lambda/(|\epsilon_\text{h}| + W_\text{h})\Big)$. The relative contribution of hot and cold electrons to the low temperature $\gamma$ at the $H \approx H_c$ is discussed further in the SI.

The observed behaviors of the specific heat and transport data shown in Fig. \ref{fig:phase} are entirely reproduced by 
the model of dominant cc $\to$ ch scattering if $T_\text{tr} \sim |\epsilon_\text{h}| + W_\text{h}$ collapses towards zero at the critical field. This collapse is cut off near the critical field by the onset of spin ordering at around 1K, as shown in Fig. \ref{fig:phase}. Therefore the width of the peak in the density of states need not strictly go to zero.

The data suggests that it is predominantly the scale $|\epsilon_\text{h}|$ that drives the collapse of $T_\text{tr}$ toward critical field. We recalled above that a peak in the specific heat is seen at a temperature $T_\text{pk}$. This Schottky anomaly-like peak in $\gamma(T)$ is quantitatively explained at zero field by the peak in the density of states seen in ARPES  \cite{PhysRevLett.101.026407}. As the temperature is raised more states in the hot region become accessible and hence the specific heat coefficient increases until the temperature crosses $\sim \max(|\epsilon_\text{h}|,W_\text{h})$, beyond which the specific heat starts to decrease (and become dominated by the cold electrons).
The simultaneous collapse of $T_\text{pk}$ and $T_\text{tr}$ as the critical field is approached is therefore naturally explained by $|\epsilon_\text{h}| \to 0$. Such behavior is consistent with a recent band structure analysis \cite{toapp}. As noted above, the essential requirement on the width $W_\text{h}$ of the peak is that it should be of order 1 kelvin at the critical field, so that the MFL behavior can persist all the way down to the onset of the low temperature ordered phase.

As $\epsilon_\text{h} \to 0$, the density of states at the chemical potential increases. This leads to an increase in the specific heat coefficient at low temperature, tying the collapse of $T_\text{tr}$ to the low temperature mass enhancement we described above.

In the scenario we have outlined, non-degenerate hot fermions are important even at low temperatures at the critical field. In this case, as noted above, elastic ch $\to$ ch scattering gives an additional contribution to the residual, temperature-independent resistivity. A strong peak in the residual resistivity at the critical field is indeed seen in the data \cite{Grigera329}. Furthermore, the Lorenz ratio at low temperatures shows increased elastic scattering at the critical field \cite{PhysRevLett.97.067005}.

Our discussion has been purely in terms of the fermionic band structure. Direct contributions to transport and thermodynamics from quantum critical collective modes is not necessary to explain the data, as considered in \cite{rostA,Mackenzie2012quantum}. Nonetheless, we mention two ways in which such bosonic physics could additionally be present. Firstly, quantum critical fluctuations of an overdamped bosonic order parameter with $z_B = 2$ in two dimensions would contribute to a logarithmic specific heat at low temperatures \cite{PhysRevB.48.7183}. Secondly, order parameter fluctuations provide an additional scattering mechanism.
A different source of $T$-linear scattering will be called for at the lowest temperatures if $W_\text{h}$ does not in fact collapse down to around 1K at the critical field (as we assumed above).

\subsection{Scattering into a divergent density of states}
\label{sec:general}

Taking a step back from Sr$_3$Ru$_2$O$_7$, nondegenerate fermions will be present in a localized region of the Brillouin zone whenever a divergence in the density of states occurs at the chemical potential, such as at a van Hove point. In general, a dispersion with the scaling $\epsilon_{\text{h}, \vec k} \sim k^z$ (this need not be a minimum in $k$, e.g. at a van Hove point $\epsilon_{\text{h}, \vec k} \sim k_x^2 - k_y^2$) leads to a density of states
\be\label{eq:dos}
\nu_\text{h}(\epsilon) \sim \epsilon^{d/z-1} \,,
\ee
which is divergent for $z>d$ (for $z=d$ there can be a logarithmic divergence).  While this section shares with our earlier discussion the importance of nondegenerate fermions, it describes a different scenario. The MFL discussed above arises at temperatures above a sharp peak in the density of states. The physics discussed here instead arises at temperature within a broader, power law or logarithmic, peak that is located at the Fermi energy. The density of states of the hot electrons diverges while $W_\text{h}$ remains finite.

The lifetime of cold fermions due to cc $\to$ ch scattering can again be computed from [\ref{eq:Sigma}] and [\ref{eq:Pi}] above. The energies are now $\omega \sim \Omega \sim \Omega' \sim T \ll E_{F\text{c}}$. The dispersion of the cold electrons can be linearized about their Fermi surfaces, as previously. The hot electrons obey
\be
\text{Im} \, G^R_\text{h}(\omega, \vec{k}) = \frac{1}{\omega} \left(\frac{W_\text{h}}{\omega}\right)^{\eta} G\left(\frac{\omega}{k^z} \frac{k_{F\text{h}}^z}{W_\text{h}} \right) \,,
\ee
for some scaling function $G$ and, for generality, allowing for an anomalous dimension $\eta$ for the hot electrons. Here $k_{F\text{h}}$ is a characteristic momentum scale of the hot fermions. This anomalous dimension further shifts the exponent in the density of states [\ref{eq:dos}], so that $\nu_\text{h}(\epsilon) \sim \int d^dk \, \text{Im} \, G^R_\text{h}(\epsilon, \vec{k}) \sim \epsilon^{d/z-1-\eta}$. The simplest case of free electrons is $\text{Im} \, G^R_\text{h}(\omega, \vec{k}) = \pi \delta(\omega - k^z \, W_\text{h}/k_{F\text{h}}^z)$. The scattering kinematics is similar to the cases considered previously, and is described in detail in the SI. The resulting scattering rate follows from scaling: of the two momentum integrals in [\ref{eq:Sigma}] and [\ref{eq:Pi}], one is over the hot region in the Brillouin zone, with $|\vec{k}'+\vec{q}| \sim T^{1/z}$, while the other is over the cold Fermi surface. The constraint that $\vec{k},\vec{k}-\vec{q}$ and $\vec{k}'$ must be near the Fermi surface means this latter integral is over a region of size $k_{F\text{c}}^{d-2}$. All told, we obtain the decay rate
\be\label{eq:Gamma2}
\Gamma_\text{c} \sim \l^2 \frac{k_{F\text{c}}^{d-2}}{v_{F\text{c}}^2} T \int_{-T}^T d\epsilon \, \nu_\text{h}(\epsilon) \sim \l^2  \frac{k_{F\text{h}}^d k_{F\text{c}}^{d-2}}{v_{F\text{c}}^2} \frac{k_B T}{\hbar} \left(\frac{k_B T}{W_\text{h}}\right)^{d/z-\eta}\,.
\ee
The scattering rate [\ref{eq:Gamma2}] is the number of states in the peak that are thermally accessible. In the final term we re-inserted factors of $\hbar$ and $k_B$.

There can be logarithmic corrections to [\ref{eq:Gamma2}]. For example, 
a van Hove point in two dimensions has $z=2$ with the dispersion $\omega \sim k_x^2 - k_y^2 = k^2 \cos(2\theta)$. This leads to a logarithmic divergence in the integrals over $q'$ in [\ref{eq:Pi}] at $\theta \approx \pm \pi/4$. Assuming the anomalous dimension $\eta$ is small, we obtain $\Gamma_\text{c} \sim T^2 \log (W_\text{h}/T)$. Another interesting general case is a multi-critical point with $z=4$ in two dimensions \cite{toapp,Yuan2019}. As long as the anomalous dimension $\eta$ is small, this gives $\Gamma_\text{c} \sim T^{3/2}$.
Both of these two scalings of $\Gamma_\text{c}$ with temperature have been obtained previously in \cite{PhysRevB.53.11344} from a variational Boltzmann equation computation. Band structure effects on transport are also considered in \cite{PhysRevB.88.245128,Herman2019}. In our discussion above, the scattering mechanism responsible for the resistivity is physically transparent.

\subsection{Scattering into the van Hove point in Sr$_2$RuO$_4$}

Under uniaxial pressure, Sr$_2$RuO$_4$ undergoes a Lifshitz transition, traversing a van Hove singularity at a critical compressive strain $\vep_\text{vH}$ \cite{Steppkeeaaf9398}. Resistivity measurements on very pure samples show a characteristic strange metal `fan' emanating from a critical strain at zero temperature~\cite{PhysRevLett.120.076602}. Similar behavior has been found upon traversing the van Hove point by chemical doping~\cite{Kikugawa2004,Shen2007} or by epitaxial strain~\cite{Burganov2016}.  Outside the fan, the resistivity $\rho \sim T^2$. In the uniaxial strain experiment the resistivity near the critical strain fits $\rho \sim T^2 \log (T_\text{vH}/T)$, with $T_\text{vH}=230$K, to excellent accuracy over a range of 40 times the residual resistivity \cite{PhysRevLett.120.076602}.
This is precisely the behavior obtained above for cc $\to$ ch scattering near a van Hove singularity. The single-particle scattering rate scales as $T$ due to ch$\to$ch proceeses~\cite{Pattnaik1992}.

At high temperatures, the resistivity of unstrained Sr$_2$RuO$_4$ gradually crosses over from quadratic to linear temperature dependence. The linear regime onsets about $600$K~\cite{Tyler1998}. It is interesting to note that this temperature is roughly comparable to the scale $T_\text{vH}$ mentioned above. This conceivably suggests that the high temperature $T$-linear behavior might also be due to cc $\to$ ch scattering into the band containing the van Hove point, although the `hot' band is not small in this case.

\subsection{Outlook}

We have shown that the strange metal phenomenology of two strontium ruthenates can be understood from the cc $\to$ ch scattering of cold electrons into small regions of hot, non-degenerate electrons. This simple process is a fermionic analogue of the well-established $T$-linear scattering rate that arises from scattering off classical bosons. It is distinct, however, from scattering off a classical fermion, which would correspond to ch $\to$ ch.\footnote{In SDW transitions, scattering off composite operators at a hot spot has been argued to lead to `lukewarm' fermions over the entire Fermi surface \cite{PhysRevB.84.125115, PhysRevB.89.155126,PhysRevB.90.045105}. This is analogous to ch $\to$ ch scattering in our framework. In particular, the fact that the scattering is small angle means that it does not strongly impact transport. It may be interesting to look at cc $\to$ ch scattering in those models.}

In Sr$_3$Ru$_2$O$_7$ in particular, the proposed dominance of cc $\to$ ch scattering successfully explains quantitative properties of dc and optical charge transport and specific heat. It will be interesting to see if more subtle observables such as the Hall coefficient, which shows a dramatic change in behavior across $T_\text{tr}$ \cite{PERRY20001469,PhysRevB.63.174435}, can also be understood from this perspective. The thermopower is also a potentially sensitive probe of the presence of non-degenerate electrons.

Our description of Sr$_3$Ru$_2$O$_7$ is structured around the presence of a peak in the density of states at zero field that collapses towards the Fermi energy as $H \to H_\text{c}$. These statements are well-grounded experimentally. The most speculative aspect of our discussion has been the assumption that the width of this peak becomes small (of order 1 kelvin) at the critical field. This is necessary within our framework to explain how the $T$-linear resistivity and $T \log 1/T$ specific heat are able to extend down to the lowest temperatures. Direct experimental confirmation of the narrowing of the peak as it moves towards the Fermi energy is needed to fully justify (or falsify) this aspect of our description. Existing scanning tunneling microscopy (STM) data as a function of field does not clearly show the peak moving towards the Fermi energy \cite{PhysRevLett.99.057208} and is therefore difficult to square with several aspects of the phenomenology.

Turning to other families of strange metals, heavy fermion systems contain narrow peaks in the density of states. A dichotomy between hot and cold fermions may be at work in those cases also, and hence they  are candidates for the approach we have taken here. For more general strange metals, however, the following caveat should be kept in mind. While controlled quantum Monte Carlo modelling of e.g. density wave or Ising nematic criticality in metals clearly shows distinct hot and cold regions of the Fermi surface \cite{qmc}, the hot regions do not necessarily lead to peaks in the density of states because the spectral weight of the hot fermions can vanish in tandem with a divergent mass \cite{PhysRevLett.117.097002}. It remains to be understood how variants of the cc $\to$ ch scattering mechanism identified in this work could underpin the strange metallic physics arising in those cases.


\subsection{Data availability}
The data and codes used in this work are available upon request.

\acknow{We have had helpful discussions with Chandra Varma, Andrew Mackenzie, Andrey Chubukov and Jan Zaanen. This work is supported by the Department of Energy, Office of Basic Energy Sciences, under Contract No. DEAC02-76SF00515. EB acknowledges financial support from the European Research Council (ERC) under grant HQMAT (grant no. 817799), the Israel-US Binational Science Foundation (BSF), and the hospitality of the Aspen Center for Physics, supported by NSF grant PHY-1607611, where part of this work was done. CHM is supported by an NSF graduate  fellowship. Collaborative work for this paper was made possible by the Nordita program 
``Bounding Transport and Chaos in Condensed Matter and Holography'', which was additionally funded by ICAM and Vetenskapradet.}

\showacknow 


\bibliography{hot}

\begin{thebibliography}{10}

\bibitem{sachdevkeimer}
Sachdev S, Keimer B (2011) Quantum criticality.
\newblock {\em Physics Today} 64:29.

\bibitem{Grigera329}
Grigera SA, et~al. (2001) {Magnetic Field-Tuned Quantum Criticality in the
  Metallic Ruthenate Sr$_3$Ru$_2$O$_7$}.
\newblock {\em Science} 294(5541):329--332.

\bibitem{PhysRevLett.101.026407}
Tamai A, et~al. (2008) {Fermi Surface and van Hove Singularities in the
  Itinerant Metamagnet ${\mathrm{Sr}}_{3}{\mathrm{Ru}}_{2}{\mathrm{O}}_{7}$}.
\newblock {\em Phys. Rev. Lett.} 101(2):026407.

\bibitem{PhysRevB.51.9253}
Hlubina R, Rice TM (1995) Resistivity as a function of temperature for models
  with hot spots on the fermi surface.
\newblock {\em Phys. Rev. B} 51(14):9253--9260.

\bibitem{PhysRevLett.82.4280}
Rosch A (1999) Interplay of disorder and spin fluctuations in the resistivity
  near a quantum critical point.
\newblock {\em Phys. Rev. Lett.} 82(21):4280--4283.

\bibitem{varma1}
Varma CM (1994) Theoretical framework for the normal state of copper oxide
  metals in {\em Strongly Correlated Electronic Systems, the Los Alamos
  Symposium 1991}, eds.{} Hebard AF, et~al.
\newblock (Addison-Wesley, Reading MA), pp. 573--603.

\bibitem{varma2}
Ruckenstein AE, Varma CM (1991) A theory of marginal fermi-liquids.
\newblock {\em Physica C: Superconductivity} 185-189:134 -- 140.

\bibitem{PhysRevLett.88.076602}
Capogna L, et~al. (2002) {Sensitivity to Disorder of the Metallic State in the
  Ruthenates}.
\newblock {\em Phys. Rev. Lett.} 88(7):076602.

\bibitem{Bruin804}
Bruin JAN, Sakai H, Perry RS, Mackenzie AP (2013) {Similarity of Scattering
  Rates in Metals Showing T-Linear Resistivity}.
\newblock {\em Science} 339(6121):804--807.

\bibitem{Rost16549}
Rost AW, et~al. (2011) {Thermodynamics of phase formation in the quantum
  critical metal Sr$_3$Ru$_2$O$_7$}.
\newblock {\em Proceedings of the National Academy of Sciences}
  108(40):16549--16553.

\bibitem{PhysRevB.78.155132}
Mirri C, et~al. (2008) {Anisotropic optical conductivity of
  ${\text{Sr}}_{3}{\text{Ru}}_{2}{\text{O}}_{7}$}.
\newblock {\em Phys. Rev. B} 78(15):155132.

\bibitem{Rost1360}
Rost AW, Perry RS, Mercure JF, Mackenzie AP, Grigera SA (2009) {Entropy
  Landscape of Phase Formation Associated with Quantum Criticality in
  Sr$_3$Ru$_2$O$_7$}.
\newblock {\em Science} 325(5946):1360--1363.

\bibitem{PhysRevLett.120.076602}
Barber ME, Gibbs AS, Maeno Y, Mackenzie AP, Hicks CW (2018) {Resistivity in the
  Vicinity of a van Hove Singularity: ${\mathrm{Sr}}_{2}{\mathrm{RuO}}_{4}$
  under Uniaxial Pressure}.
\newblock {\em Phys. Rev. Lett.} 120(7):076602.

\bibitem{PhysRevLett.95.127001}
Kitagawa K, et~al. (2005) {Metamagnetic Quantum Criticality Revealed by
  $^{17}\mathrm{O}\mathrm{\text{\ensuremath{-}}}\mathrm{NMR}$ in the Itinerant
  Metamagnet ${\mathrm{Sr}}_{3}{\mathrm{Ru}}_{2}{\mathrm{O}}_{7}$}.
\newblock {\em Phys. Rev. Lett.} 95(12):127001.

\bibitem{PhysRevB.97.115101}
Sun D, Rost AW, Perry RS, Mackenzie AP, Brando M (2018) {Low temperature
  thermodynamic investigation of the phase diagram of
  ${\mathrm{Sr}}_{3}{\mathrm{Ru}}_{2}{\mathrm{O}}_{7}$}.
\newblock {\em Phys. Rev. B} 97(11):115101.

\bibitem{Lester2015}
Lester C, et~al. (2015) {Field-tunable spin-density-wave phases in
  Sr$_3$Ru$_2$O$_7$}.
\newblock {\em Nature Materials} 14:373.

\bibitem{AllanARPES}
Allan MP, et~al. (2013) {Formation of heavy d-electron quasiparticles in
  Sr$_3$Ru$_2$O$_7$}.
\newblock {\em New Journal of Physics} 15(6):063029.

\bibitem{PhysRevB.81.235103}
Mercure JF, et~al. (2010) {Quantum oscillations near the metamagnetic
  transition in ${\text{Sr}}_{3}{\text{Ru}}_{2}{\text{O}}_{7}$}.
\newblock {\em Phys. Rev. B} 81(23):235103.

\bibitem{CAVA1995141}
Cava R, et~al. (1995) {Sr$_2$RuO$_4$ $\cdot$ 0.25 CO$_2$ and the Synthesis and
  Elementary Properties of Sr$_3$Ru$_2$O$_7$}.
\newblock {\em Journal of Solid State Chemistry} 116(1):141 -- 145.

\bibitem{PhysRevB.62.R6089}
Ikeda SI, Maeno Y, Nakatsuji S, Kosaka M, Uwatoko Y (2000) {Ground state in
  ${\mathrm{Sr}}_{3}{\mathrm{Ru}}_{2}{\mathrm{O}}_{7}:$ Fermi liquid close to a
  ferromagnetic instability}.
\newblock {\em Phys. Rev. B} 62(10):R6089--R6092.

\bibitem{PhysRevLett.63.1996}
Varma CM, Littlewood PB, Schmitt-Rink S, Abrahams E, Ruckenstein AE (1989)
  {Phenomenology of the normal state of Cu-O high-temperature superconductors}.
\newblock {\em Phys. Rev. Lett.} 63(18):1996--1999.

\bibitem{Mahan2000many}
Mahan G (2000) Many-body physics.
\newblock {\em Springer}.

\bibitem{Zaanen2004}
Zaanen J (2004) Why the temperature is high.
\newblock {\em Nature} 430:512.

\bibitem{KADOWAKI1986507}
Kadowaki K, Woods S (1986) Universal relationship of the resistivity and
  specific heat in heavy-fermion compounds.
\newblock {\em Solid State Communications} 58(8):507 -- 509.

\bibitem{PhysRevLett.86.2661}
Perry RS, et~al. (2001) {Metamagnetism and Critical Fluctuations in High
  Quality Single Crystals of the Bilayer Ruthenate
  ${\mathrm{Sr}}_{3}{\mathrm{Ru}}_{2}{O}_{7}$}.
\newblock {\em Phys. Rev. Lett.} 86(12):2661--2664.

\bibitem{PhysRevLett.88.217204}
Millis AJ, Schofield AJ, Lonzarich GG, Grigera SA (2002) Metamagnetic quantum
  criticality in metals.
\newblock {\em Phys. Rev. Lett.} 88(21):217204.

\bibitem{PhysRevLett.116.226402}
Tokiwa Y, Mchalwat M, Perry RS, Gegenwart P (2016) {Multiple Metamagnetic
  Quantum Criticality in ${\mathrm{Sr}}_{3}{\mathrm{Ru}}_{2}{\mathrm{O}}_{7}$}.
\newblock {\em Phys. Rev. Lett.} 116(22):226402.

\bibitem{toapp}
Efremov DV, et~al. (2018) Multicritical fermi surface topological transitions.

\bibitem{PhysRevLett.97.067005}
Ronning F, et~al. (2006) Thermal conductivity in the vicinity of the quantum
  critical end point in ${\mathrm{sr}}_{3}{\mathrm{ru}}_{2}{\mathrm{o}}_{7}$.
\newblock {\em Phys. Rev. Lett.} 97(6):067005.

\bibitem{rostA}
Rost AW, et~al. (2010) {Power law specific heat divergence in
  Sr$_3$Ru$_2$O$_7$}.
\newblock {\em Physica Status Solidi (b)} 247(3):513--515.

\bibitem{Mackenzie2012quantum}
Mackenzie A, Bruin J, Borzi R, Rost A, Grigera S (2012) {Quantum criticality
  and the formation of a putative electronic liquid crystal in
  Sr$_3$Ru$_2$O$_7$}.
\newblock {\em Physica C: Superconductivity} 481:207 -- 214.
\newblock Stripes and Electronic Liquid Crystals in Strongly Correlated
  Materials.

\bibitem{PhysRevB.48.7183}
Millis AJ (1993) Effect of a nonzero temperature on quantum critical points in
  itinerant fermion systems.
\newblock {\em Phys. Rev. B} 48(10):7183--7196.

\bibitem{Yuan2019}
{Yuan} NFQ, {Isobe} H, {Fu} L (2019) {Magic of high order van Hove
  singularity}.

\bibitem{PhysRevB.53.11344}
Hlubina R (1996) Effect of impurities on the transport properties in the van
  hove scenario.
\newblock {\em Phys. Rev. B} 53(17):11344--11347.

\bibitem{PhysRevB.88.245128}
Buhmann JM (2013) Unconventional scaling of resistivity in two-dimensional
  fermi liquids.
\newblock {\em Phys. Rev. B} 88(24):245128.

\bibitem{Herman2019}
Herman Fcv, Buhmann J, Fischer MH, Sigrist M (2019) Deviation from fermi-liquid
  transport behavior in the vicinity of a van hove singularity.
\newblock {\em Phys. Rev. B} 99(18):184107.

\bibitem{Steppkeeaaf9398}
Steppke A, et~al. (2017) {Strong peak in T$_c$ of Sr$_2$RuO$_4$ under uniaxial
  pressure}.
\newblock {\em Science} 355(6321):9398.

\bibitem{Kikugawa2004}
Kikugawa N, Bergemann C, Mackenzie AP, Maeno Y (2004) {Band-selective
  modification of the magnetic fluctuations in
  ${\mathrm{Sr}}_{2}{\mathrm{RuO}}_{4}$: A study of substitution effects}.
\newblock {\em Phys. Rev. B} 70(13):134520.

\bibitem{Shen2007}
Shen KM, et~al. (2007) {Evolution of the Fermi Surface and Quasiparticle
  Renormalization through a van Hove Singularity in
  ${\mathrm{Sr}}_{2\ensuremath{-}y}{\mathrm{La}}_{y}{\mathrm{RuO}}_{4}$}.
\newblock {\em Phys. Rev. Lett.} 99(18):187001.

\bibitem{Burganov2016}
Burganov B, et~al. (2016) {Strain Control of Fermiology and Many-Body
  Interactions in Two-Dimensional Ruthenates}.
\newblock {\em Phys. Rev. Lett.} 116(19):197003.

\bibitem{Pattnaik1992}
Pattnaik PC, Kane CL, Newns DM, Tsuei CC (1992) Evidence for the van hove
  scenario in high-temperature superconductivity from quasiparticle-lifetime
  broadening.
\newblock {\em Phys. Rev. B} 45(10):5714--5717.

\bibitem{Tyler1998}
Tyler AW, Mackenzie AP, NishiZaki S, Maeno Y (1998) {High-temperature
  resistivity of ${\mathrm{Sr}}_{2}{\mathrm{RuO}}_{4}:$ Bad metallic transport
  in a good metal}.
\newblock {\em Phys. Rev. B} 58(16):R10107--R10110.

\bibitem{PhysRevB.84.125115}
Hartnoll SA, Hofman DM, Metlitski MA, Sachdev S (2011) Quantum critical
  response at the onset of spin-density-wave order in two-dimensional metals.
\newblock {\em Phys. Rev. B} 84(12):125115.

\bibitem{PhysRevB.89.155126}
Chubukov AV, Maslov DL, Yudson VI (2014) Optical conductivity of a
  two-dimensional metal at the onset of spin-density-wave order.
\newblock {\em Phys. Rev. B} 89(15):155126.

\bibitem{PhysRevB.90.045105}
Abrahams E, Schmalian J, W\"olfle P (2014) Strong-coupling theory of
  heavy-fermion criticality.
\newblock {\em Phys. Rev. B} 90(4):045105.

\bibitem{PERRY20001469}
Perry R, et~al. (2000) {Hall effect of Sr$_3$Ru$_2$O$_7$}.
\newblock {\em Physica B: Cond. Mat.} 284-288:1469 -- 1470.

\bibitem{PhysRevB.63.174435}
Liu Y, et~al. (2001) {Electrical transport properties of single-crystal
  ${\mathrm{Sr}}_{3}{\mathrm{Ru}}_{2}{\mathrm{O}}_{7}:$ The possible existence
  of an antiferromagnetic instability at low temperatures}.
\newblock {\em Phys. Rev. B} 63(17):174435.

\bibitem{PhysRevLett.99.057208}
Iwaya K, et~al. (2007) {Local Tunneling Spectroscopy across a Metamagnetic
  Critical Point in the Bilayer Ruthenate
  ${\mathrm{Sr}}_{3}{\mathrm{Ru}}_{2}{\mathrm{O}}_{7}$}.
\newblock {\em Phys. Rev. Lett.} 99(5):057208.

\bibitem{qmc}
Berg E, Lederer S, Schattner Y, Trebst S (2019) Monte carlo studies of quantum
  critical metals.
\newblock {\em Annual Review of Condensed Matter Physics} 10(1):63--84.

\bibitem{PhysRevLett.117.097002}
Schattner Y, Gerlach MH, Trebst S, Berg E (2016) Competing orders in a nearly
  antiferromagnetic metal.
\newblock {\em Phys. Rev. Lett.} 117(9):097002.

\end{thebibliography}

\newpage


\renewcommand{\thefigure}{S\arabic{figure}}
\renewcommand{\theequation}{S\arabic{equation}}
\setcounter{figure}{0}
\setcounter{equation}{0}

\section*{Supplementary Material}

\subsection{Kinematics of general scattering processes}

In the main text we focused on cc $\to$ ch scattering. Here we firstly discuss in a little more detail why other scattering processes do not contribute as substantially to transport:

\begin{enumerate}

\item cc $\rightarrow$ cc processes give a scattering rate scaling as $T^2$. Due to the relatively low density of states of the cold electrons, these contributions will be seen to be sub-dominant compared to processes involving both c and h electrons.

\item  The ch $\rightarrow$ ch contribution to the transport lifetime is suppressed because such processes either involve a small momentum transfer (if the h electron scatters within the same hot pocket), or they involve very specific momenta corresponding to the inter-hot pocket distance (if the h electron scatters between different hot pockets). Scattering processes of the latter kind can only occur between special momenta on the cold Fermi surface, that are connected to each other by an inter-hot pocket distance. We expect these regions of the cold Fermi surface to be `short-circuited' by other regions that do not allow such resonant ch $\rightarrow$ ch scattering. Similar considerations apply for ch $\rightarrow$ hh scattering processes.

\item cc $\rightarrow$ hh processes of the Cooper type (i.e., scattering a pair of cold electrons with opposite momenta to a pair of hot electrons at hot pockets $\pm \vec{k}_h$) can occur over the entire cold Fermi surface. However, such processes do not modify the current (or any odd moment of the electronic distribution function), and thus they do not contribute to transport. Further, cc $\rightarrow$ hh processes with large center of mass momentum of the incoming particles can only occur on a small portion of the cold Fermi surface.

\end{enumerate}

Furthermore, we should note that if the cold Fermi surface is sufficiently large, the cc $\to$ ch scattering will include umklapp processes which degrade momentum. Otherwise, the total momentum of electrons in the hot regions must be rapidly degraded by some other process, so that there are no subtleties associated with overall momentum conservation. For the specific case of Sr$_3$Ru$_2$O$_7$ it is straightforward to convince oneself that all of the cold electrons are able to participate in umklapp scattering into a hot pocket. In the following Fig. \ref{fig:umklapp} we illustrate two such scatterings, in which the bands seen at the Fermi energy scatter into the hot region. The band structure of Sr$_3$Ru$_2$O$_7$ will be discussed in more detail in Sec.~\ref{sec:phase} below, when we consider the available phase space for scattering.

\begin{figure}[h]
\begin{center}
\includegraphics[width=0.45\textwidth]{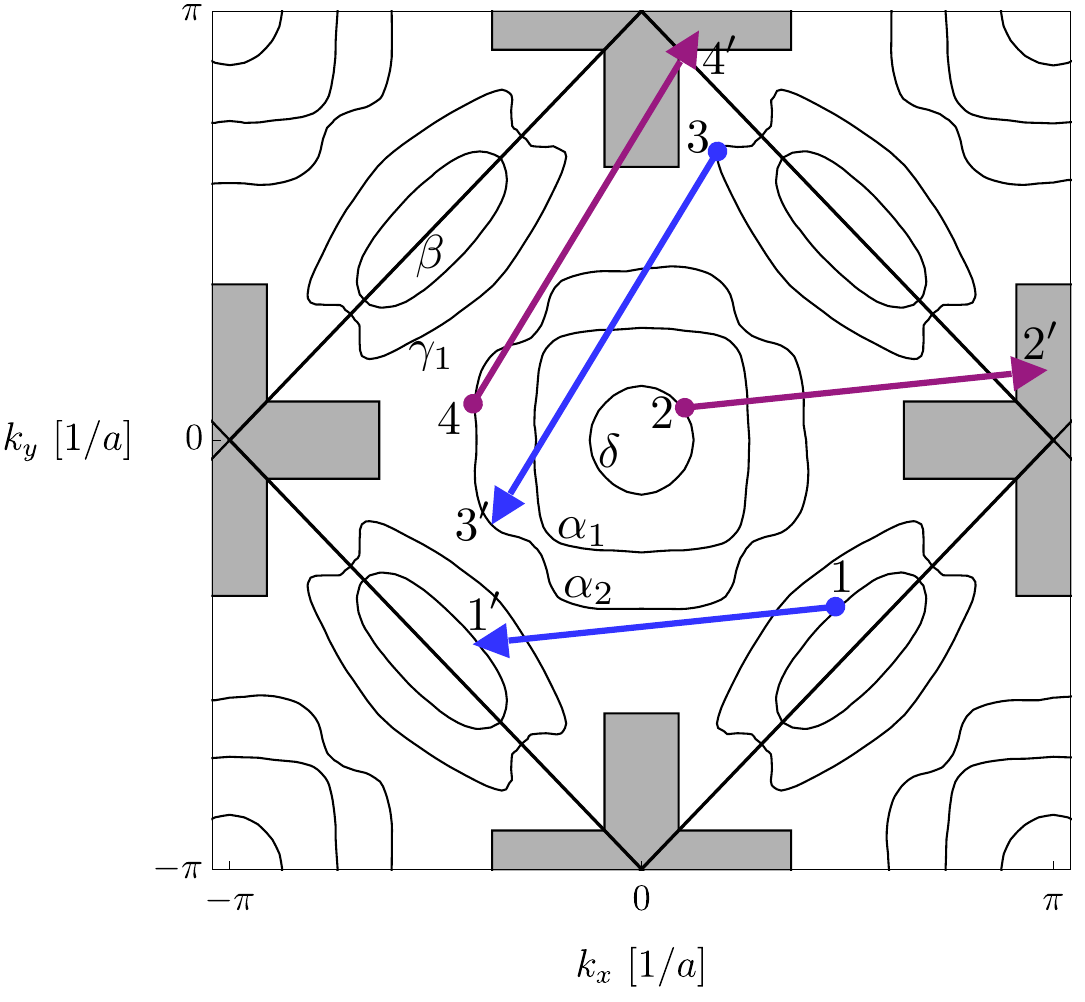}\caption{Two examples of umklapp scattering into the hot region in Sr$_3$Ru$_2$O$_7$.
The shaded area represents the hot regions of the Brillouin zone (see Sec.~\ref{sec:phase} below). The momentum states that participate in the scattering processes are labeled as $(1,2)\rightarrow(1’,2’)$ and $(3,4)\rightarrow(3’,4’)$.
}
\label{fig:umklapp}
\end{center}
\end{figure}

\subsection{Comments on the Planckian scattering rate}

In the main text we estimate the scattering rate to be `Planckian', $\Gamma \sim k_B T/\hbar$. This conclusion came from data for the Fermi momenta and velocities and a dimensional analysis estimate of the interaction strength $\lambda$. Here we make two further comments:

\begin{enumerate}

\item The value of $\lambda$ actually drops out of the scattering rate if one extrapolates the theory to large values of $\lambda$. In this limit the single-particle inverse lifetime, defined as $\Gamma_\text{c}= \Sigma_\text{c}''(0)/(1 + \partial \Sigma_\text{c}'/\partial\omega|_{\omega=0})$, never exceeds the Planckian value of $k_BT/\hbar$ times an $O(1)$ number. This is since for large $\lambda$ the real part of the MFL self-energy, $\Sigma'_c(\omega)\sim \lambda^2 (k_{F\text{h}}/v_{F\text{c}})^2 \omega \log(E_{F\text{c}}/\omega)$, dominates over the bare $\omega$ term in the inverse propagator of the cold fermions, and as a result $\lambda$ drops out of $\Gamma_{\text{c}}$.

\item The Planckian scattering rate is due, among other things, to the fact that the sum of areas of the hot pockets is comparable to a momentum space scale set by the cold Fermi surfaces: $\sum_i k_{F\text{h},i}^2 \sim k_{F\text{c}}^2$. This conclusion can be reached by an independent argument involving the contribution of hot and cold fermions to the specific heat. At $H=H_c$, the logarithmic dependence of $\gamma(T)$ extends down to the SDW ordering temperature, $T_\text{SDW} \sim 1$K. One might worry that the large entropy of the degenerate hot fermions should instead make a dominant contribution to the specific heat in this regime. Indeed, at lower temperatures in the data $\gamma(T)$ rises above the extrapolated logarithmic dependence, see Ref. (10) in the main text. We ascribe the enhancement of $\gamma(T)$ right below $T_\text{SDW}$ to the hot electrons, which are becoming either degenerate or gapped by the SDW order. From the magnitude of the increase in $\gamma$ at $T_{\text{SDW}}$  one can estimate that the contribution of the hot electrons at low temperatures, $\gamma_{\text{hot}}\sim k_{F\text{h}}^2/W_{\text{h}}$ is comparable to the cold contribution $\Delta \gamma_{\text{cold}} \sim \Delta m_{\star\text{c}} \sim m_{\star\text{c}}$. This latter order of magnitude estimate (from the data) implies that $\sum_i k_{F\text{h},i}^2 \sim k_{F\text{c}}^2$, which coincides with the estimate given above.

\end{enumerate}

\subsection{Kinematics of cc $\rightarrow$ ch scattering} 
\label{app:MFL}

\begin{figure}[h]
\begin{center}
\includegraphics[width=0.45\textwidth]{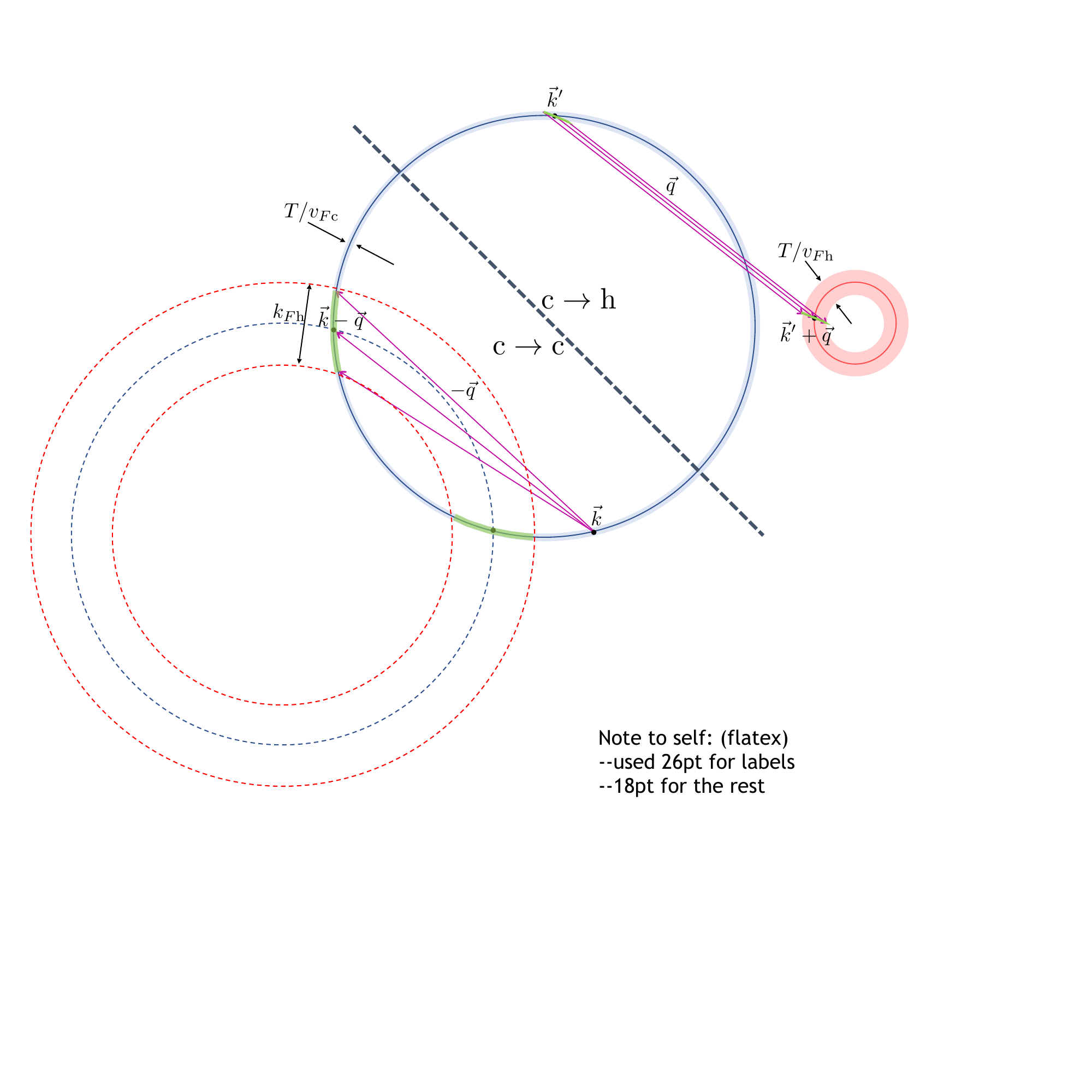}
\caption{Kinematics for cc $\rightarrow$ ch scattering. Top right half shows the phase space available to c $\rightarrow$ h scattering for a fixed momentum transfer $\vec{q}$ and energy transfer $\Omega=0$ as representative cases. The green arcs, connected by $\vec{q}$, depict the phase space available for $\vec{k'}$; they have length $\sim \frac{T}{v_{F\text{h}}}$ when the hot fermions are degenerate, and $\sim k_{F\text{h}}$ when the entire hot spot can be excited. Lower left half shows the kinematics for scattering the incoming electron, c $\rightarrow$ c. The blue dashed circle --- representing a fictitious, momentum-shifted cold Fermi surface --- denotes the space of  $\vec{k}-\vec{q}$ permitted by momentum conservation so that $\vec{q}$ can connect the cold Fermi surface to precisely the center of the hot spot. For $\vec{q}$ to connect the cold Fermi surface to any point in the hot spot, $\vec{k}-\vec{q}$ must lie in the dashed red annulus of width $\sim k_{F\text{h}}$. Energetics further restrict $\vec{k}-\vec{q}$ to be within $\frac{T}{v_{F\text{c}}}$ of the cold Fermi surface.}
\label{fig:kinematics_ccch}
\end{center}
\end{figure}

Here we derive the scattering rates [5] and [6] in the main text, respectively $\Gamma_\text{c}\sim T^1$ and $\Gamma_\text{c}\sim T^2$, which are independent of the detailed nature of the hot band structure. Our calculation here is based on a geometrical approach summarized in Fig. \ref{fig:kinematics_ccch} to determine the phase space available for cc $\rightarrow$ ch scattering. The focus here is on the Brillouin zone geometry of scattering. The following section gives a more detailed derivation, encompassing all cases we have considered.

We work with dimensions $d=2$, but generalize our arguments to any dimension $d$ at the end. The hot and cold electrons spectral functions are taken to be $\text{Im} G^R_\text{h/c}(\omega,\vec k) = \pi \delta(\omega - \epsilon_\text{h/c}(\vec k))$.
The self energy as defined in [1] and [2] in the main text can be simplified greatly by
approximating the Fermi-Dirac functions as $f(\omega)\sim 1$ for $\omega\lesssim T$ and $0$ otherwise, so that $\Omega'$ and $\Omega$ are both restricted to a range of $-T$ to $T$. After this approximation, performing the integral over $\Omega'$ in (2) tells us that
\begin{equation}\label{eq:Pi_geometric}\Pi_\text{ch}''(\Omega,\vec{q}\,)\sim \frac{1}{b(\Omega)}
\int_{\epsilon_{\text{h},\vec{k}'}\in(-T,T)}d^2k'\, \delta(\Omega-\epsilon_{\text{c},\vec{k}'+\vec{q}}+\epsilon_{\text{h},\vec{k}'}) \,.
\end{equation}
This measures the phase space available to $\vec{k'}$, assuming a fixed energy difference $\Omega$ and momentum difference $\vec{q}$ between the cold and hot electron. On the upper right half of Fig. \ref{fig:kinematics_ccch} the phase space available to $\vec{k'}$ is denoted by the one-dimensional green arcs. This arc has a length  $\frac{T}{v_{F\text{h}}}$ at low temperatures near the bottom of the peak in the density of states, where the hot electrons are degenerate. At temperatures above the peak, $T > |\epsilon_\text{h}|+W_\text{h}$, the entire hot spot is accessible and the arc length saturates at $k_{F\text{h}}$. Defining $k'_{\parallel}$ and $k'_{\perp}$ to run parallel and perpendicular to the arc, respectively, the polarization bubble becomes:
\begin{eqnarray}\label{eq:Pi_final}
\Pi_\text{ch}''(\Omega,\vec{q}\,) & \sim 
\frac{1}{b(\Omega)}\int_{-\min(\frac{T}{v_{F\text{h}}}
,\,k_{F\text{h}})}^{\min(\frac{T}{v_{F\text{h}}},\,k_{F\text{h}})} dk'_{\parallel}
\int dk'_{\perp} \frac{\delta(k'_{\perp})}{|\nabla_{k'}(\epsilon_{\text{c},\vec{k}'+\vec{q}}-\epsilon_{\text{h},\vec{k}'})|} \nonumber \\ 
& \sim \frac{1}{b(\Omega)}\frac{1}{v_{F\text{c}}}\min(\frac{T}{v_{F\text{h}}},k_{F\text{h}}).
\end{eqnarray}
This result is valid only when $|\Omega|\lesssim T$ and $\vec{q}$ can connect the hot spot and cold Fermi surface; otherwise, $\Pi''_{\text{ch}}(\Omega,\vec{q}\,)\approx 0$. At intermediate temperatures thermal broadening can lead to a $T^{1-\alpha}$ temperature dependence of the accessible hot electron density of states, that is more general than in [\ref{eq:Pi_final}]. This regime is captured by the calculation in
the following section. The gradient $|\nabla_{k'} (\epsilon_{\text{c},\vec{k}'+\vec{q}}-\epsilon_{\text{h},\vec{k}'})|$ scales like $v_{F\text{c}}\gg v_{F\text{h}}$.

Inserting the result for the polarization bubble [\ref{eq:Pi_final}] into [1] and assuming $\omega\in(-T,T)$,
\begin{equation}
\Sigma_\text{c}''(\omega,\vec{k})\sim \frac{\lambda^2}{v_{F\text{c}}}\min(\frac{T}{v_{F\text{h}}},k_{F\text{h}})
\int_{\substack{\text{connects} \\ \text{c}\rightarrow\text{h}}} d^2 q \int_{-T}^{T} d\Omega~ \delta(\omega-\Omega-\epsilon_{\text{c},\vec{k}-\vec{q}})\,.
\end{equation}
Performing the integral over $\Omega$, we find that $\Sigma_{c}''$ scales like the phase space available to $\vec{q}$:
\begin{equation}
\Sigma_\text{c}''(\omega,\vec{k})\sim \frac{\lambda^2}{v_{F\text{c}}}\min(\frac{T}{v_{F\text{h}}},k_{F\text{h}}) \int_{\substack{\text{connects c $\rightarrow$ h} \\ \epsilon_{\text{c},\vec{k}-\vec{q}}\,\in(-T,T)}}d^2q \,.
\end{equation}
In the lower left half of Fig. \ref{fig:kinematics_ccch} the first restriction in the integral over $\vec{q}$ forces $\vec{k}-\vec{q}$ to lie within momentum $k_{F\text{h}}$ of the blue dashed circle. The second restriction requires $\vec{k}-\vec{q}$ to lie within $\frac{T}{v_{F\text{c}}}$ of the cold Fermi surface. The two regions denoted by these restrictions intersect in the green arcs of length $\sim k_{F\text{h}}$ and width $\sim \frac{T}{v_{F\text{c}}}$, so that the phase space available to $\vec{q}$ goes like $\frac{T}{v_{F\text{c}}}k_{F\text{h}}$.
It follows that
\begin{equation}
\Sigma_\text{c}''(\Omega,\vec{q}\,)\sim
\lambda^2 \min(\frac{T}{v_{F\text{h}}},k_{F\text{h}})\frac{k_{F\text{h}}}{v_{F\text{c}}}\frac{T}{v_{F\text{c}}}\,,
\end{equation}
thereby reproducing [5] and [6] in the main text.

It is straightforward to generalize these geometric arguments to general dimension $d$. The momentum $\vec{k}'$ has phase space $\sim k_{F\text{h}}^{d-2}\min(k_{F\text{h}},\frac{T}{v_{F\text{h}}})$ because the one-dimensional arc in the upper right of Fig. \ref{fig:kinematics_ccch} becomes a $(d-1)$-dimensional surface. Similarly, the momentum $\vec{q}$ has phase space $\sim k_{F\text{c}}^{d-2}k_{F\text{h}}\times \frac{T}{v_{F\text{c}}}$. The self energy for general dimension $d$ is therefore
\begin{equation}
\Sigma_\text{c}''(\vec{q},\Omega)
\sim \lambda^2 \frac{k_{F\text{c}}^{d-2}k_{F\text{h}}^{d-1}}{v_{F\text{c}}^2} \min(k_{F\text{h}},\frac{T}{v_{F\text{h}}})\,T\,.
\end{equation}

\subsection{Detailed derivation for a general hot electron dispersion}
\label{app:gen}

\begin{figure}[h]
\begin{center}
\includegraphics[width=0.45\textwidth]{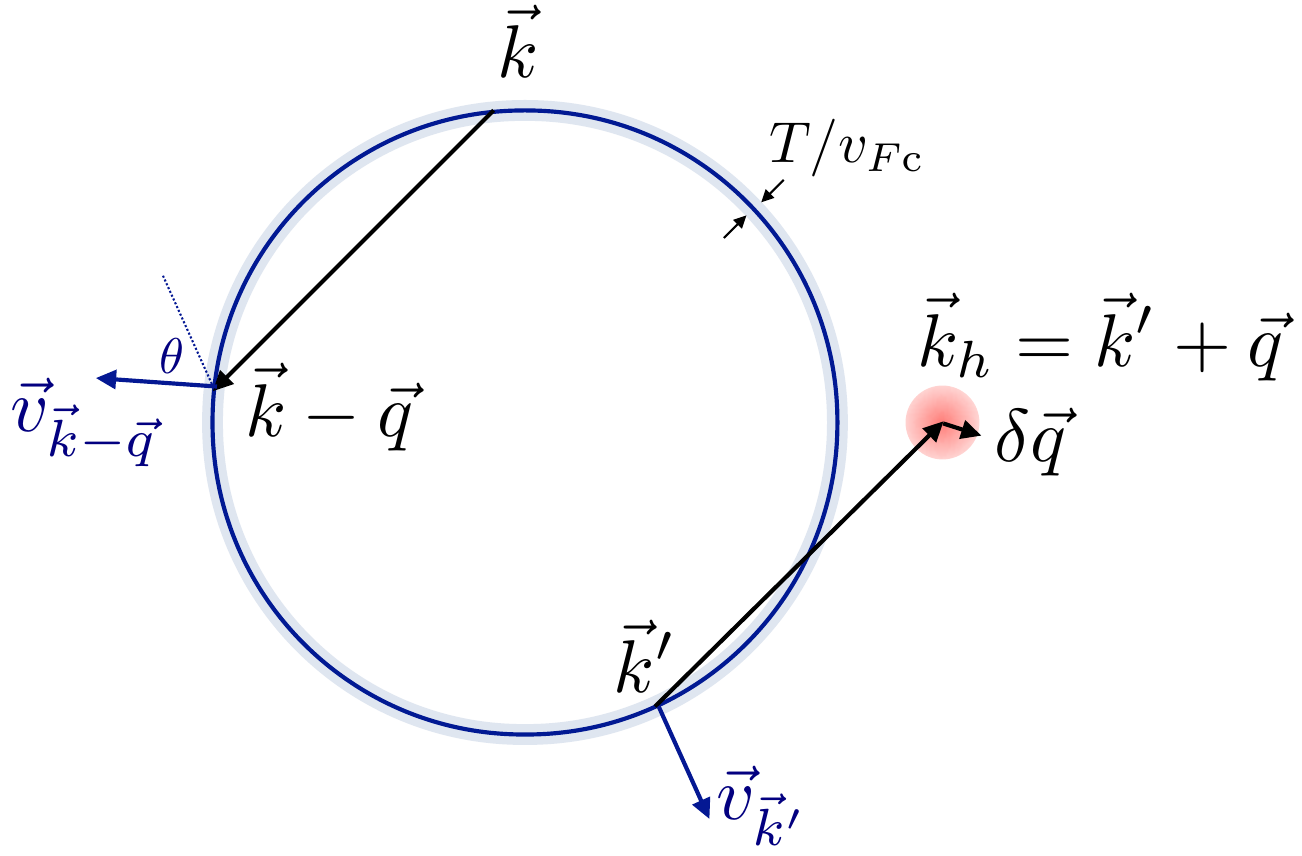}\caption{Geometry for the scattering rate of an electron at $\vec{k}$ due to a cc$\rightarrow$ch process.}
\label{fig:general}
\end{center}
\end{figure}

Here, we give a detailed derivation of the scattering rate due to
the cc $\rightarrow$ ch process for a general hot electron density
of states. The dispersions of the cold and hot electrons
are $\epsilon_{\mathrm{c},\vec{k}}$ and $\epsilon_{\mathrm{h},\vec{k}}$, respectively,
and consider the decay rate of an electron at momentum $\vec{k}$
on the cold Fermi surface ($\epsilon_{\mathrm{c}\vec{,k}}=0$), as a function
of temperature. The geometry of the scattering process is illustrated
in Fig. \ref{fig:general}.
We choose $\vec{k}'$ on the cold Fermi
surface such that $\vec{k}-\vec{q}$ (where $\vec{q}=\vec{k}_{h}-\vec{k}'$)
is also on the cold Fermi surface. By Fermi's golden rule, the decay
rate of the electron at $\vec{k}$ is given by\footnote{There are in fact two distinct scattering processes that need to be
summed over, related by an exchange of the two outgoing electrons, each with
its own matrix element. The cc$\rightarrow$ch scattering amplitude
$\lambda$ includes both processes.}$^,$\footnote{For the total scattering rate
out of $\vec{k}$, we should add also the ch$\rightarrow$cc process;
the analysis of this process is similar to that of cc$\rightarrow$ch,
and gives the same parameteric temperature dependence.}
\begin{align}
\Gamma_{\mathrm{c}}(\vec{k},T) & =2\pi\lambda^{2}\int\frac{d^{2}\delta qd^{2}\delta k'}{(2\pi)^{4}}\delta\left(\epsilon_{\mathrm{c},\vec{k}'+\delta\vec{k}'}-\epsilon_{\mathrm{c},\vec{k}-\vec{q}-\delta\vec{q}}-\epsilon_{\mathrm{h},\vec{k}_{h}+\delta\vec{k}'+\delta\vec{q}}\right)\nonumber \\
 & \times f(\epsilon_{\mathrm{c},\vec{k}'+\delta\vec{k}'})\left[1-f(\epsilon_{\mathrm{c},\vec{k}-\vec{q}-\delta\vec{q}})\right]\left[1-f(\epsilon_{\mathrm{h},\vec{k}_{h}+\delta\vec{k}'+\delta\vec{q}})\right].\label{eq:Gamma_c}
\end{align}

We have approximated, here, that the hot electrons are free so that $\text{Im}\,G_\text{h}$ is a delta function. Linearizing the dispersion near the cold Fermi surface,
\begin{align}
\epsilon_{\mathrm{c},\vec{k}-\vec{q}-\delta\vec{q}} & =-\vec{v}_{k-q}\cdot\delta\vec{q}=-v_{F\text{c}}(\delta q_{\perp}\cos\theta+\delta q_{\parallel}\sin\theta)\nonumber \\
\epsilon_{\mathrm{c},\vec{k}'+\delta\vec{k}'} & =\vec{v}_{k'}\cdot\delta\vec{k}'=v_{F\text{c}}\delta k_{\perp}',\label{eq:linearize}
\end{align}
where $\perp$ and $\parallel$ denote the components of either $\delta\vec{k}'$
or $\delta\vec{q}$ perpendicular and parallel to the cold Fermi surface
at $\vec{k}'$, respectively. $\theta$ is the angle between $-\vec{v}_{\vec{k}'}$
and $\vec{v}_{\vec{k}-\vec{q}}\,$. For simplicity, we have assumed
a circular cold Fermi surface.

It is useful to perform a change of variables,
\begin{align}
\delta\tilde{q}_{\perp} & =\delta q_{\perp}\cos\theta+\delta q_{\parallel}\sin\theta,\nonumber \\
\delta\tilde{q}_{\parallel} & =-\delta q_{\perp}\sin\theta+\delta q_{\parallel}\cos\theta.\label{eq:change}
\end{align}
Using [\ref{eq:linearize},\ref{eq:change}], Eq. [\ref{eq:Gamma_c}]
becomes
\begin{align}
\Gamma_{\mathrm{c}}(\vec{k},T) & =2\pi\lambda^{2}\int\frac{d^{2}\delta\tilde{q}d^{2}\delta k'}{(2\pi)^{4}}\delta\left(v_{F\text{c}}\delta k_{\perp}'+v_{F\text{c}}\delta\tilde{q}_{\perp}-\epsilon_{\mathrm{h},\vec{k}_{h}+\delta\vec{k}'+\delta\vec{q}}\right)\nonumber \\
 & \times f(v_{F\text{c}}\delta k_{\perp}')[1-f(-v_{F\text{c}}\delta\tilde{q}_{\perp})][1-f(v_{F\text{c}}\delta k_{\perp}'+v_{F\text{c}}\delta\tilde{q}_{\perp})].\label{eq:Gamma_c2}
\end{align}
Let us define the generalized hot density of states
\begin{equation}
\tilde{\nu}_{\mathrm{h}}(\varepsilon)=\int\frac{d\left(\delta k'_{\parallel}\right)d\left(\delta\tilde{q}_{\parallel}\right)}{\left(2\pi\right)^{2}}\delta\left(\varepsilon-\epsilon_{\mathrm{h},\vec{k}_{h}+\delta\vec{k}'+\delta\vec{q}}\right).
\end{equation}
Finally, we change variables from $(\delta k'_{\parallel},\delta\tilde{q}{}_{\parallel})$
to 
\be
(k_{1},k_{2})=(-\delta\tilde{q}{}_{\parallel}\sin\theta,\delta\tilde{q}{}_{\parallel}\cos\theta+\delta k'_{\parallel}) + (\delta \tilde{q}_\perp\cos\theta + \delta k_\perp,\delta \tilde{q}_\perp\sin\theta).\ee
This gives
\begin{equation}
\tilde{\nu}_{\mathrm{h}}(\varepsilon)=\frac{2}{|\sin2\theta|}\int\frac{dk_{1}dk_{2}}{\left(2\pi\right)^{2}}\delta\left(\varepsilon-\epsilon_{\mathrm{h},\vec{k}_{h}+(k_{1},k_{2})}\right)=\frac{2}{|\sin2\theta|}\nu_{\mathrm{h}}(\varepsilon),\label{eq:nu_gen}
\end{equation}
where $\nu_{\mathrm{h}}(\varepsilon)$ is the hot density of states. The factor
$2/|\sin2\theta|$ originates from the Jacobian; we have assumed that
$\sin2\theta\ne0$. At points where $\sin2\theta=0$, the cold dispersion
cannot be linearized, and a more careful treatment is needed. However,
since this occurs only at a discrete set of points, the transport
properties are not expected to be affected. 

Inserting [\ref{eq:nu_gen}] in [\ref{eq:Gamma_c2}], we arrive at the result:
\begin{equation}
\Gamma_\text{c}(\vec{k},T)\sim \lambda^2 \frac{T}{v_{F\text{c}}^2}\int_{-T}^{T} d\epsilon\,\nu_\text{h}(\epsilon),\label{eq:gamC}
\end{equation}
We consider the divergent hot density of states $\nu_\text{h}(\epsilon)\sim (\epsilon-\epsilon_\text{h})^{-\alpha}$ [or $\nu_\text{h}(\epsilon)\sim \ln\left(\frac{|\epsilon-\epsilon_\text{h}|}{W_\text{h}}\right)$, corresponding to the case of a van Hove singularity in $d=2$] within an energy range of $W_\text{h}$. At low temperatures $T< |\epsilon_\text{h}|$ where the peak in $\nu_\text{h}(\epsilon)$ is not excited, $\Gamma_c\sim T^2$. At intermediate temperatures $|\epsilon_\text{h}|<T<W_\text{h}$, the $\epsilon$ integral gives $T^{1-\alpha}$ [or $T\ln(T/W_\text{h})$, in the van Hove case], so that $\Gamma_c\sim T^{2-\alpha}$  [or $T^2\ln(T/W_\text{h})$], respectively. The former is consistent with Eq.~[9] in the main text, with $1+\eta-d/z=\alpha$. At high temperatures $W_\text{h}<T$, the universal $\Gamma_\text{c}\sim T^{1}$ is recovered.

\begin{figure}[t]
\begin{center}
\includegraphics[width=0.35\textwidth]{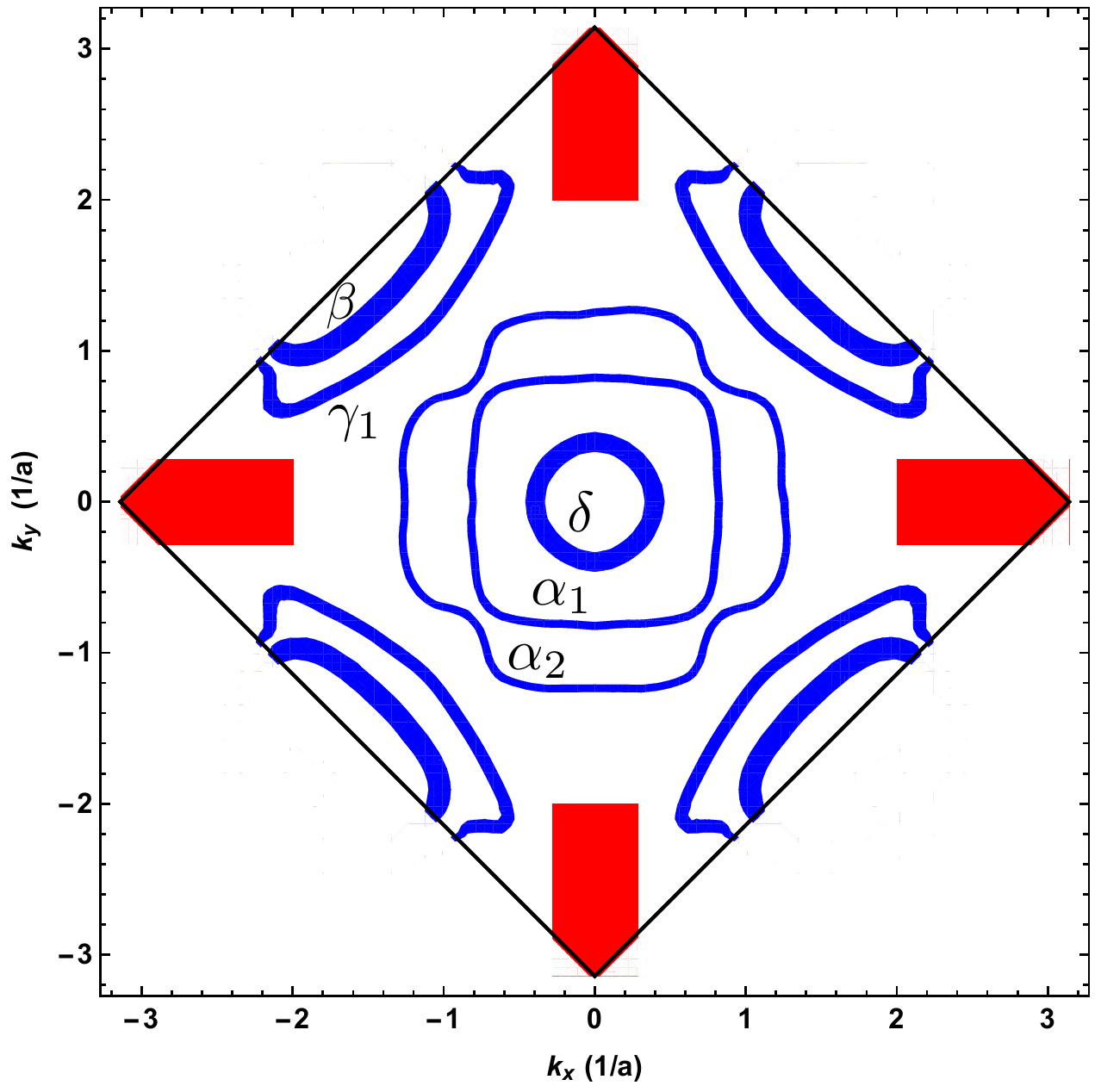}\caption{Regions of the Brillouin Zone occupied by cold fermions at a temperature of 20 K (in blue), as well as the approximate region occupied by hot fermions (in red). The cold regions are defined by energies between $-20$ K and $20$ K, determined from treating each band velocity (taken from Refs.~\cite{Bruin804,PhysRevB.81.235103}) as isotropic. Fermi surface contours and hot regions from Ref.~\cite{PhysRevLett.101.026407}. The lattice parameter is taken to be $a = 3.89$ A.}
\label{fig:BZ}
\end{center}
\end{figure}

\subsection{Onset of T-linear scattering}
\label{app:Tlin}

\begin{figure}[h]
\begin{center}
\includegraphics[width=0.4\textwidth]{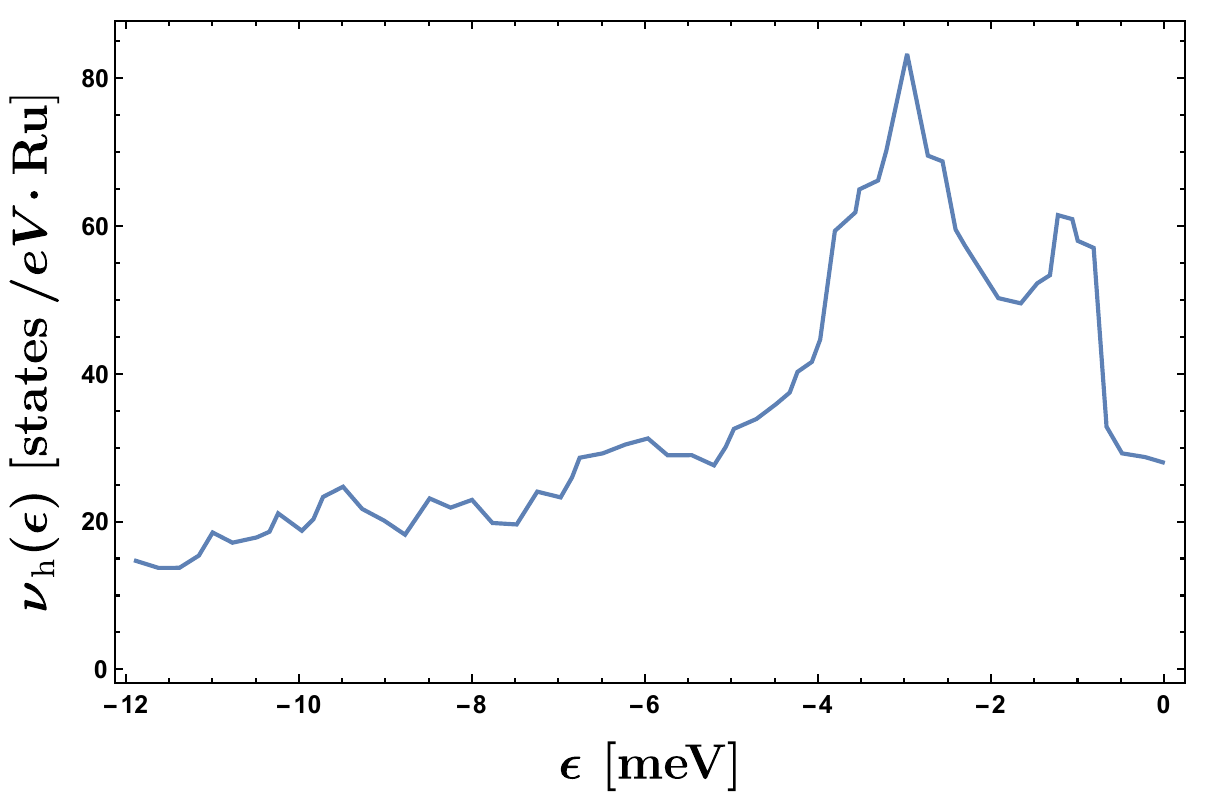}
\includegraphics[width=0.4\textwidth]{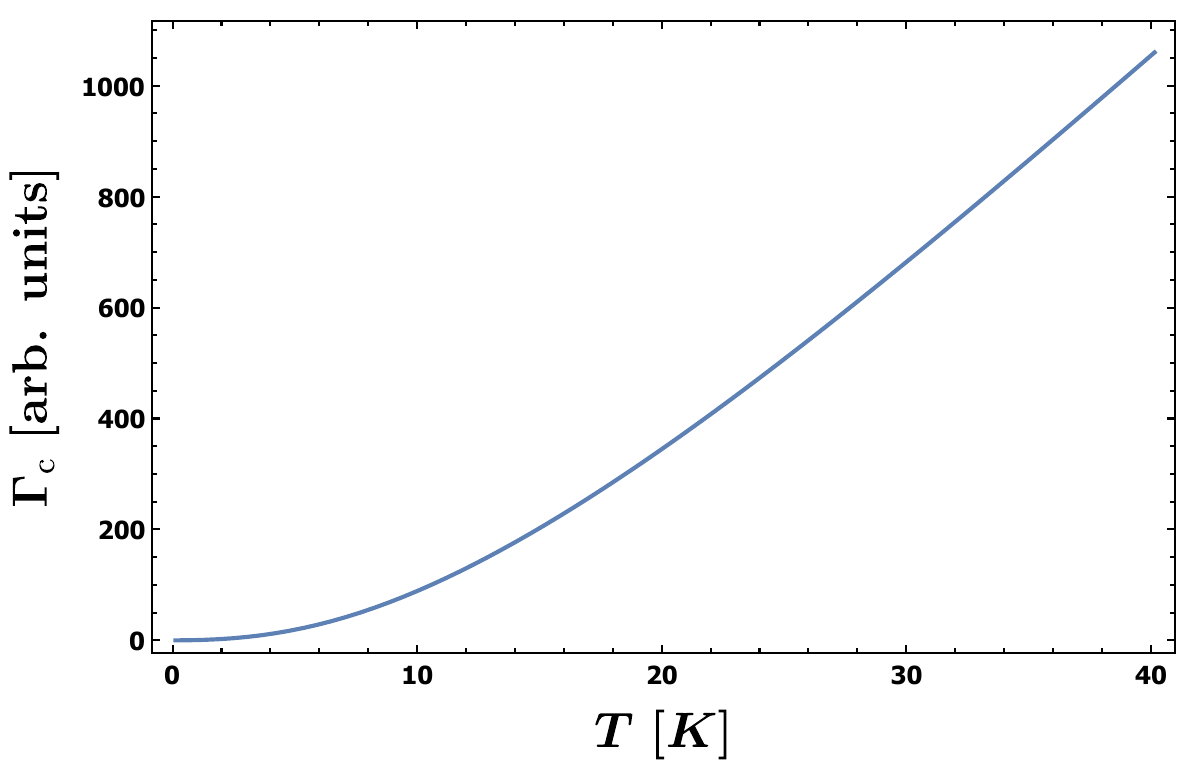}\caption{(Left) observed density of states for the hot electron peak (from Ref.~\cite{PhysRevLett.101.026407}). (Right) Scattering rate [\ref{eq:care}] calculated from hot density of states, due to cc $\rightarrow$ ch scattering in Sr$_3$Ru$_2$O$_7$.}
\label{fig:Gam}
\end{center}
\end{figure}

We can evaluate [\ref{eq:Gamma_c2}] more carefully to get a more precise version of Eq. [\ref{eq:gamC}]. Integrating over the momenta at fixed hot fermion energy $\epsilon$ gives
\be
\Gamma_\text{c} \propto \frac{\lambda^2}{v_{F\text{c}}^2} \int d\epsilon \frac{\nu_\text{h}(\epsilon) \, \epsilon}{\sinh(\epsilon/k_B T)} \,.\label{eq:care}
\ee
The hot electron density of states $\nu_\text{h}(\epsilon)$ for Sr$_3$Ru$_2$O$_7$ has been determined from ARPES measurements in Ref.~\cite{PhysRevLett.101.026407}, and is reproduced in Fig.~\ref{fig:Gam} (left). Using that data in the integral [\ref{eq:care}] we obtain the temperature-dependent scattering rate shown in 
Fig.~\ref{fig:Gam} (right).
The figure exhibits a crossover from $\Gamma_\text{c} \sim T^2$ to $\Gamma_\text{c} \sim T$ at the scale $T_\text{tr} \approx 20$K, in agreement with the transport data discussed in the main text. It is interesting to note that $T$-linear scattering onsets before the temperature reaches the center of the peak in the density of states, which is at around 2 -- 4 meV below the chemical potential.

\subsection{Accessible phase space for hot and cold scattering in Sr$_3$Ru$_2$O$_7$}
\label{sec:phase}

The cc $\rightarrow$ ch scattering we have focused on competes with cc $\rightarrow$ cc scattering. The latter leads to a conventional $T^2$ scattering rate. In order for a strong $T$-linear scattering to be visible, the phase space for cc $\rightarrow$ ch scattering should be comparable or greater than that for cc $\rightarrow$ cc scattering (assuming that the various scatterings have comparable interaction strengths). In the main text we explained that this condition can be met fairly naturally because the region of the Brillouin Zone occupied by the hot fermions is of order $k_\text{h}^2$, while a cold fermion occupies
a region of size $k_\text{c} T/v_{F\text{c}}$ and $T/v_{F\text{c}}$ is small for temperatures well below the Fermi energy.

Sr$_3$Ru$_2$O$_7$ has a complicated band structure, which in turn complicates the estimation of the available scattering phase space as a function of temperature. We will use the ARPES data from Refs.~\cite{PhysRevLett.101.026407,AllanARPES}, together with Fermi velocities extracted from quantum oscillations in Refs.~\cite{Bruin804,PhysRevB.81.235103}. There are five cold bands at the chemical potential. However, the bands have a nontrivial structure already at low energies of a few meV from the chemical potential. This means that the well-characterized band structure at the chemical potential alone is insufficient to determine the scattering phase space at temperatures that are well within the observed $T$-linear regime.
For example, the $\delta$ band has a depth of only about 8 meV below the chemical potential. The Fermi velocity of this band at the chemical potential is small, only a factor of 2 greater than that of the hot fermions. At relatively low temperatures, then, it is possible that this band should be considered as `hot'.

We have therefore estimated the available phase space at $T = 20$ K, the lowest temperature showing $T$-linear transport in zero field. At this temperature it should be reasonable to use the band structure at the chemical potential. Fig. \ref{fig:BZ} shows the region of the Brillouin Zone occupied by thermally broadened cold bands (in blue) at this temperature, together with the approximate area of the hot, almost flat band (in red). The thickness of the blue lines is $\frac{2 k_B T}{v_F}$, such that thicker cold bands have a lower Fermi velocity. The red regions denote the approximate area of the Brillouin zone where there is a nearly-flat band (to within $\pm 1$meV) centered at an energy of about $3$meV below the Fermi energy~\cite{PhysRevLett.101.026407} (this is the region that gives rise to the peak in the DOS shown in Fig.~\ref{fig:Gam} above).
In comparing the total area occupied in the Brillouin Zone, one must account for the multiplicities discussed in Refs.~\cite{Bruin804,PhysRevB.81.235103}. For the bands $(\delta, \alpha_1, \alpha_2, \gamma_1, \beta)$  this involves multiplying the area shown in Fig. \ref{fig:BZ} by $(2,1,1,2,1)$ in addition to an overall factor of 2 for the spin. The area of the hot region shown should be multiplied by a factor of 4 (including the spin multiplicity). The thermally broadened cold bands are found to occupy a phase space of roughly $(1.3, 0.61, 0.84, 3.8, 2.0)/a^2$, summing up to $9.0/a^2$, while the hot regions have a total phase space of about $10/a^2$. Therefore, indeed, the hot electron phase competes with and slightly dominates over the cold electron phase at this temperature.


\end{document}